\documentclass[reprint,amsmath,amssymb,aps]{revtex4-1}
\usepackage[colorlinks,citecolor=blue,linkcolor=blue,anchorcolor=blue,filecolor=blue,
urlcolor=blue]{hyperref}
\usepackage{graphicx}% Include figure files
\usepackage{ulem}
\usepackage{dcolumn}% Align table columns on decimal point
\usepackage{bm}% bold math
\usepackage{todonotes}

\begin{document}

\title{Dark matter component in hadronic models with short-range correlations}

\author{O. Louren\c{c}o, T. Frederico, and M. Dutra}  
\affiliation{Departamento de F\'isica, Instituto Tecnol\'ogico de Aeron\'autica, DCTA, 12228-900, S\~ao Jos\'e dos Campos, SP, Brazil}

\date{\today}

\begin{abstract}
A relativistic mean field hadronic model with a dark matter~(DM) particle coupled to nucleons 
including short-range correlations~(SRC) is applied to study neutron stars (NS). The lightest 
neutralino is chosen as the dark particle candidate, which interacts with nucleons by the exchange 
of Higgs bosons. A detailed thermodynamical analysis shows that the contribution of the DM fermions 
to the energy density of the matter composed by these particles and nucleons is completely dominated 
by the DM kinetic terms. The model reproduces satisfactorily the constraints on the mass-radius 
diagram obtained from the  analysis of the combined data from the NICER mission, LIGO and 
Virgo collaborations, and mass measurements from radio observations. We show that the SRC balance 
the reduction of the neutron star mass due to the~DM component, and because of that the model is 
able to present more massive~NS. We also present a study of the effect, in the NS mass-radius 
profiles, of the uncertainties in some bulk parameters related to the hadronic sector. We find that 
it is possible to generate parametrizations, with DM content, compatible with the recent 
astrophysical constraints and with the uncertainty in the symmetry energy slope obtained from the 
results reported by the updated Lead Radius EXperiment~(PREX-2). 
\end{abstract}

% \PACS{x}

\maketitle

\section{Introduction} 

A complete understanding of the fundamental physics involving hadronic matter is still a challenge 
for both, theoretical and experimental research. For instance, in the case of matter composed by 
strongly interacting particles at zero temperature and high density regime, the results provided by 
the available hadronic models can be tested against constraints coming from astrophysical 
observational data. In particular, compact objects, such as neutron stars~(NS), are considered 
natural laboratories to test matter under extreme conditions (very dense systems). Therefore, dense 
stellar matter becomes an important source of information on the forces in nature. 

In that direction, a large amount of data has also been provided by the so-called multi-messenger 
astronomy era~\cite{era}, initiated by the detection of gravitational waves signal produced by the 
collision of two black holes~\cite{bholes1,bholes2,bholes3}, and the latter on by two NS in a merger 
mechanism~\cite{merger}. Another relevant source of information regarding features of astrophysical 
systems is NASA's Neutron star Interior Composition Explorer~(NICER), namely, an X-ray telescope 
installed on the International Space Station~\cite{nicer}. From the NICER data related to the 
massive millisecond pulsars, such as the PSR J0030+0451 
one~\cite{psr0451-1,psr0451-2,psr0451-3,psr0451-4,psr0451-5,psr0451-6}, it is possible to estimate 
masses and radii of~NS. Recently, a new round of measurements of the NICER mission is available, now 
regarding the PSR J0740+6620 pulsar~\cite{arx78,arx79,arx80,arx81}. 

Another important component that can directly affect the description of astrophysical and 
cosmological systems is dark matter (DM), whose fundamental nature is still unknown (see 
Ref.~\cite{dmrev} and references therein for an overview). The existence of this kind of (dark) 
particle is due to measurements made by F.~Zwicky~\cite{zwicky}, who verified a high dispersion of 
the velocities of galaxies in the Coma cluster, a phenomenon that can not be explained exclusively 
through ordinary matter. A similar study was also performed by J.~Oort~\cite{oort}. In his 
investigation, Oort verified too high velocities of stars in the solar neighborhood. Once again, the 
results of this observation are not completely understood if  only luminous matter is considered. 
Other studies on the rotation curves of different galaxies were also performed, with the same result 
inferred, namely, the existence of matter other than the known visible matter. Furthermore, the use 
of gravitational lensing methods~\cite{lensing,lensing2} leads to a visible mass of approximately 
10\% to 20\% for the total mass of galaxy clusters, the result that also supports the emergence of 
extra matter in such structures. Other relevant evidence for the presence of dark matter comes from 
measurements of the anisotropy of cosmic microwave background (CMB)~\cite{cmb1,cmb2,cmb3,cmb4,cmb5}. 
The current understanding predicts that 27\% of the universe is made of dark matter, 68\% of dark 
energy (the main component that explains the accelerated expansion of the universe), and only 5\% of 
luminous matter.

Since the aforementioned evidence points out the necessity to investigate new physics, the next 
step is to understand what kind of particles compose such (dark) matter. Among the possible 
candidates, the most promising ones are the so-called Weakly Interacting Massive Particles (WIMPs), 
namely, weak-scale particles produced in the early universe as a thermal relic of the Big Bang, and 
typically with mass in the range of $1$~GeV to $100$~TeV. The predicted results of relic abundances 
are compatible with corresponding observational data if WIMPs are taken as DM 
particles~\cite{planck}. Other kinds of candidates are also considered in order to explain 
observations related to DM. For instance, the model of Asymmetric Dark Matter (ADM)~\cite{adm} is 
based on the possibility of having a matter/anti-matter asymmetry for dark particles likewise one 
observes in ordinary baryons. The masses of ADM particles, around $5$~GeV, are lower in comparison 
with the previous case. Other no-WIMP candidates can be listed here, such as the superWIMP 
particles, namely, massive particles whose interaction is much weaker than those presented by WIMPs. 
Gravitinos and axinos are examples of these superWIMP candidates. Furthermore, one also has axions, 
sterile neutrinos, WIMPzillas, supersymmetric Q-balls, mirror matter, and even black holes as viable 
candidates. Interesting discussions and details on these particles can be found in 
Refs.~\cite{cand1,cand2}, for instance.

Recently, some studies were performed in which DM models were coupled to hadronic ones and used to 
describe astrophysical systems, such as neutron stars. In the hadronic sector, there are many models 
available for the nuclear and stellar matter. As examples, we cite the nonrelativistic 
Skyrme~\cite{sky1,sky2,sky3,stoneskyrme,had4} and Gogny~\cite{gogny1,gogny7,gogny8,gognyic} 
parametrizations, and relativistic mean-field (RMF) models~\cite{rev1,rev2,rev3,had}. The first 
class was used in Refs.~\cite{eftdm1,eftdm2}, and the latter one in 
Refs.~\cite{rmfdm1,rmfdm2,rmfdm3,rmfdm4,rmfdm5,rmfdm6,rmfdm7,rmfdm8,rmfdm9}. Furthermore, other 
source of valuable information regarding the nuclear interaction comes from microscopic calculations 
based on the chiral effective field theory 
(EFT)~\cite{eft,machleidt16,whitehead20,hebeler,tews,drischler,hebeler2,drischler2,melendez,
drischler3,samma}. Here we use the RMF approach, but for the hadronic sector of the combined 
description (DM + nuclear matter) we generalize the model in order to include effects from 
nucleon-nucleon short-range correlations 
(SRC)~\cite{nature,hen2017,ye2018,Egiyan2006,Frankfurt1993,Fomin2012,Atti2015,Shneor2007,Tang2003,
Li2019,Schmookler2019,Duer2019,Ryck2019,Chen2017}. This phenomenology establishes that 
nonindependent nucleons correlate in pairs with high relative momentum as a consequence of the 
short-range part of the nuclear interaction. Besides, SRC can  also be probed by experiments such as 
those performed at the Thomas Jefferson National Accelerator Facility (JLab)~\cite{subedi2008}, in 
which the collision of very energetic incident electrons with the $^{12} \rm C$ nucleus, for 
instance, induces the removal of two correlated nucleons with large relative momentum, and in this 
case mostly neutron-proton pairs. This is a feature also observed in other target nuclei, namely, 
$^{27}\rm Al$, $^{56}\rm Fe$, and $^{208}\rm Pb$~\cite{orhen}. A direct implication of SRC is the 
change of the nucleon momentum distribution, $n(k)$, in nuclei and nuclear matter. In this case, the 
usual step function of a Fermi gas of independent nucleons is replaced by a ``High Momentum Tail'' 
(HMT) distribution, with the ``tail'' usually taken as proportional to $1/k^4$~\cite{cai,lucas}. 
Here, the same form is used for the SRC contribution to the RMF description of the DM-hadronic 
equations of state~(EOS).

In this work, we investigate the influence of  SRC's in the hadronic model coupled to DM and show 
that it is possible to generate mass-radius diagrams in agreement with recent data provided by the 
analysis of the NICER mission observations~\cite{arx78,arx79,arx80,arx81}. This analysis also 
includes gravitational wave data obtained from 
LIGO and 
Virgo collaborations~\cite{era,bholes1,bholes2,bholes3,merger,ligo1,ligo2,ligo3,ligo4,ligo5,ligo6} 
and maximum mass measurements from radio observations. We split the study into two parts. In 
Sec.~\ref{part1}, we show that the SRC balance the decreasing of the neutron star mass caused by the 
inclusion of dark matter. Furthermore, we also verify that the DM itself can be implemented in the 
system only by adding the kinetic part of the dark fermion energy. In Sec.~\ref{part2}, we study the 
consequences of varying the
bulk parameters in the hadronic sector within the accepted ranges to estimate theoretical uncertainties. In particular, for the symmetry energy slope, we take the range of $L_0=(106\pm37)$~MeV. These numbers are in agreement with results reported by the updated Lead Radius EXperiment~(PREX-2)~\cite{piekaprex2,prex2}, and also overlaps with the range obtained from the analysis of charged pions spectra~\cite{pions}.

Our paper is organized as follows. In Sec.~\ref{form} we present the Lagrangian density for the coupled \mbox{DM-hadron} model, and the main thermodynamical quantities obtained within the RMF approach with SRC included. For the DM sector, we consider the lightest neutralino interacting with hadrons by the exchange of the Higgs boson. In Sec.~\ref{res}, we present our results for the NS mass-radius diagram with the DM component and compare with the recent astrophysical data. We also check our findings against the constraints provided in Ref.~\cite{cromartie} concerning the boundaries of the neutron star mass. This section is separated into two parts, as already mentioned. Finally, our summary and concluding remarks are given in Sec.~\ref{secsummary}.

\section{ DM-hadron model with SRC} 
\label{form}

The Lagrangian density used to describe the hadronic matter addressed here, namely, nucleons and mesons, reads
\begin{align}
&\mathcal{L}_{\mbox{\tiny HAD}} = \overline{\psi}(i\gamma^\mu\partial_\mu - M_{\mbox{\tiny nuc}})\psi 
+ g_\sigma\sigma\overline{\psi}\psi 
- g_\omega\overline{\psi}\gamma^\mu\omega_\mu\psi
\nonumber \\ 
&- \frac{g_\rho}{2}\overline{\psi}\gamma^\mu\vec{\rho}_\mu\vec{\tau}\psi
+\frac{1}{2}(\partial^\mu \sigma \partial_\mu \sigma - m^2_\sigma\sigma^2)
- \frac{A}{3}\sigma^3 - \frac{B}{4}\sigma^4 
\nonumber\\
&-\frac{1}{4}F^{\mu\nu}F_{\mu\nu} 
+ \frac{1}{2}m^2_\omega\omega_\mu\omega^\mu 
+ \frac{C}{4}(g_\omega^2\omega_\mu\omega^\mu)^2 -\frac{1}{4}\vec{B}^{\mu\nu}\vec{B}_{\mu\nu} 
\nonumber \\
&+ \frac{1}{2}\alpha'_3g_\omega^2 g_\rho^2\omega_\mu\omega^\mu
\vec{\rho}_\mu\vec{\rho}^\mu + \frac{1}{2}m^2_\rho\vec{\rho}_\mu\vec{\rho}^\mu,
\label{dlag}
\end{align}
in which $\psi$ is the nucleon field whereas $\sigma$, $\omega^\mu$, and $\vec{\rho}_\mu$ are the scalar, vector, and isovector-vector fields representing, respectively, mesons $\sigma$, $\omega$, and $\rho$. Furthermore, one has $F_{\mu\nu}=\partial_\nu\omega_\mu-\partial_\mu\omega_\nu$ and $\vec{B}_{\mu\nu}=\partial_\nu\vec{\rho}_\mu-\partial_\mu\vec{\rho}_\nu$. The nucleon rest mass is $M_{\mbox{\tiny nuc}}$, and the mesons masses are $m_\sigma$, $m_\omega$, and $m_\rho$. In the first version of this model, due to J. D. Walecka~\cite{walecka}, meson self-interactions were absent, like the ones whose strengths are given by the constants $A$, $B$,  $C$, and the ones between the $\omega$ and $\vec{\rho}$, regulated by the constant $\alpha_3'$. Here we use a more sophisticated structure for the hadronic model in which these couplings are not set to zero, i. e., $A$, $B$, $C$ and $\alpha_3'$ are nonvanishing. 

Concerning the dark matter description, we assume a dark fermion represented by the Dirac field $\chi$ that interacts with nucleons through the exchange of the Higgs boson, whose mass is $m_h$. The field associated with this particular boson is denoted by $h$. At this point, we consider the dark particle candidate with mass given by $M_\chi$. Therefore, the Lagrangian density describing the total system (nucleons, mesons, and dark matter) is written as
\begin{align}
\mathcal{L} &= \overline{\chi}(i\gamma^\mu\partial_\mu - M_\chi)\chi
+ yh\overline{\chi}\chi +\frac{1}{2}(\partial^\mu h \partial_\mu h - m^2_h h^2)
\nonumber\\
&+ f\frac{M_{\mbox{\tiny nuc}}}{v}h\overline{\psi}\psi + \mathcal{L}_{\mbox{\tiny HAD}},
\label{dlagtotal}
\end{align}
in which $fM_{\mbox{\tiny nuc}}/v$ is the Higgs-nucleon coupling with $v=246$~GeV being the Higgs vacuum expectation value. The strength of the Higgs-dark particle coupling is regulated by the constant~$y$. The equations that determine nucleon, mesons, Higgs boson, and dark particle fields are obtained from the Euler-Lagrange equations. As a consequence of the mean-field approximation,  we consider all mediator fields treated as
classical ones, namely, 
\begin{eqnarray}
\sigma\rightarrow \left<\sigma\right>\equiv\sigma, \quad
\omega_\mu\rightarrow \left<\omega_\mu\right>\equiv\omega_0
\end{eqnarray}
\begin{eqnarray}
\vec{\rho}_\mu\rightarrow \left<\vec{\rho}_\mu\right>\equiv \bar{\rho}_{0(3)}, \quad
h\rightarrow \left<h\right>\equiv h,
\end{eqnarray}
one has
\begin{align}
m^2_\sigma\sigma &= g_\sigma\rho_s - A\sigma^2 - B\sigma^3 
\\
m_\omega^2\omega_0 &= g_\omega\rho - Cg_\omega(g_\omega \omega_0)^3 
- {\alpha_3}'g_\omega^2 g_\rho^2\bar{\rho}_{0(3)}^2\omega_0, 
\\
m_\rho^2\bar{\rho}_{0(3)} &= \frac{g_\rho}{2}\rho_3 
-{\alpha_3}'g_\omega^2 g_\rho^2\bar{\rho}_{0(3)}\omega_0^2, 
\\
[\gamma^\mu (&i\partial_\mu - V) - M^*]\psi = 0,
\\
m^2_hh &= y\rho_s^{\mbox{\tiny DM}} + f\frac{M_{\mbox{\tiny nuc}}}{v}\rho_s
\\
(\gamma^\mu &i\partial_\mu - M_\chi^*)\chi = 0,
\end{align}
with effective nucleon and dark effective masses given, respectively, by
\begin{align}
M^* &= M_{\mbox{\tiny nuc}} - g_\sigma\sigma - f\frac{M_{\mbox{\tiny nuc}}}{v}h
\label{mnuc}
\\
M^*_\chi &= M_\chi - yh.
\label{mneu}
\end{align}
Notice that, in principle, the Higgs boson also affects the nucleon effective mass, since $h$ is present in Eq.~(\ref{mnuc}). Furthermore, one also has $V = g_\omega\omega_0 + \frac{g_\rho}{2}\bar{\rho}_{0(3)}\tau_3$ with $\tau_3=1$ for protons and $\tau_3=-1$ for neutrons, and
\begin{align}
\rho_s &=\left<\overline{\psi}\psi\right>={\rho_s}_p+{\rho_s}_n,
\label{rhos}
\\
\rho &=\left<\overline{\psi}\gamma^0\psi\right> = \rho_p + \rho_n,
\\
\rho_3&=\left<\overline{\psi}\gamma^0{\tau}_3\psi\right> = \rho_p - \rho_n=(2y-1)\rho,
\\
\rho_s^{\mbox{\tiny DM}} &= \left<\overline{\chi}\chi\right>,
\end{align}
where
\begin{eqnarray}
\rho_s^{\mbox{\tiny DM}} &=& 
\frac{\gamma M^*_\chi}{2\pi^2}\int_0^{k_F^{\mbox{\tiny DM}}} \hspace{-0.5cm}\frac{k^2dk}{(k^2+M^{*2}_\chi)^{1/2}}.
\end{eqnarray}
The indices $p,n$ stand for protons and neutrons, respectively. The degeneracy factor is $\gamma=2$ and the proton fraction is defined as $y_p=\rho_p/\rho$, with proton/neutron densities given by $\rho_{p,n}=\gamma{k_F^3}_{p,n}/(6\pi^2)$. The quantities ${k_F}_{p,n}$ and $k_F^{\mbox{\tiny DM}}$ are the Fermi momentum associated with the nucleon (protons and neutrons), and the dark matter particle, respectively. 

\begin{figure}[!htb] 
\centering
\includegraphics[scale=0.3]{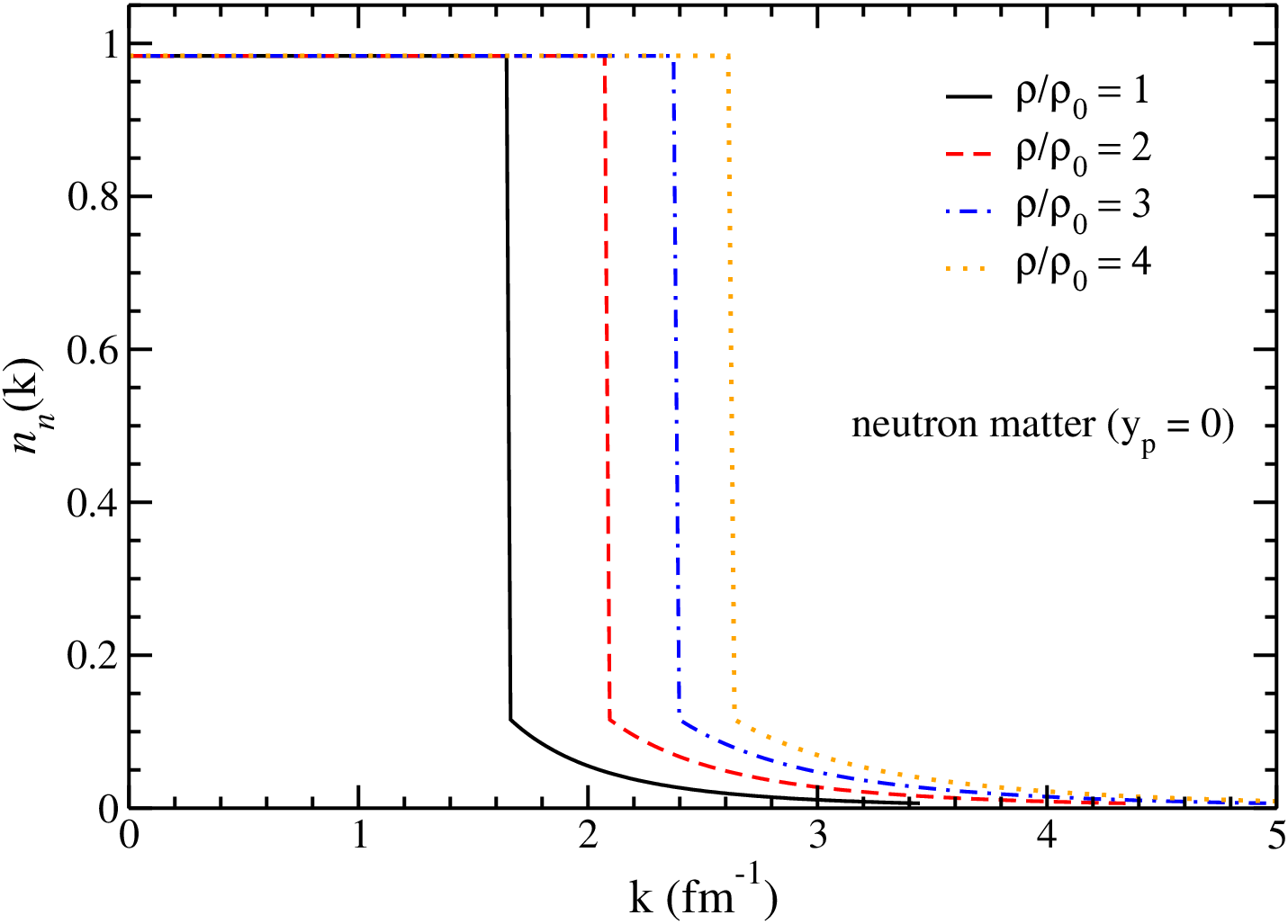}
\caption{Momentum distribution with HMT, Eq.~(\ref{eqhtm}), for neutron matter. Curves for $\rho/\rho_0=1,2,3$ and $4$.} 
\label{hmt}
\end{figure}

From the field equations of this generalized model (nucleons + dark matter), it is possible to determine the energy density and the pressure of the system. These thermodynamic quantities, obtained from the energy-momentum tensor $T^{\mu\nu}$ as $\mathcal{E}=\left<T_{00}\right>$ and $P=\left<T_{ii}\right>/3$, read
\begin{align} 
&\mathcal{E} = \frac{m_{\sigma}^{2} \sigma^{2}}{2} +\frac{A\sigma^{3}}{3} +\frac{B\sigma^{4}}{4} 
-\frac{m_{\omega}^{2} \omega_{0}^{2}}{2} - \frac{Cg_{\omega}^4\omega_{0}^4}{4}
- \frac{m_{\rho}^{2} \bar{\rho}_{0(3)}^{2}}{2} 
\nonumber\\
&+ g_{\omega} \omega_{0} \rho + \frac{g_{\rho}}{2} 
\bar{\rho}_{0(3)} \rho_{3}   -\frac{1}{2} \alpha'_3 g_{\omega}^{2} g_{\rho}^{2} \omega_{0}^{2} 
\bar{\rho}_{0(3)}^{2} + \mathcal{E}_{\mathrm{kin}}^{p} + \mathcal{E}_{\mathrm{kin}}^{n}
\nonumber\\
&+ \frac{m_h^2h^2}{2} + \mathcal{E}_{\mathrm{kin}}^{\mbox{\tiny DM}},
\label{eden}
\end{align}
and
\begin{align}
&P = -\frac{m_{\sigma}^{2} \sigma^{2}}{2} - \frac{A\sigma^{3}}{3} - \frac{B\sigma^{4}}{4} 
+ \frac{m_{\omega}^{2} \omega_{0}^{2}}{2} + \frac{Cg_{\omega}^4\omega_0^4}{4}
\nonumber\\
&+ \frac{m_{\rho}^{2} \bar{\rho}_{0(3)}^{2}}{2} + \frac{1}{2} \alpha'_3 g_{\omega}^{2} 
g_{\rho}^{2} \omega_{0}^{2} \bar{\rho}_{0(3)}^{2} + P_{\mathrm{kin}}^{p} + P_{\mathrm{kin}}^{n}
- \frac{m_h^2h^2}{2} 
\nonumber\\
&+ P_{\mathrm{kin}}^{\mbox{\tiny DM}},
\label{press}
\end{align}
in which the kinetic contributions from the dark fermion are
\begin{eqnarray}
\mathcal{E}_{\mbox{\tiny kin}}^{\mbox{\tiny DM}} &=& \frac{\gamma}{2\pi^2}\int_0^{k_F^{\mbox{\tiny DM}}}\hspace{-0.3cm}k^2(k^2+M^{*2}_\chi)^{1/2}dk,
\label{ekindm}
\\
P_{\mbox{\tiny kin}}^{\mbox{\tiny DM}} &=& 
\frac{\gamma}{6\pi^2}\int_0^{{k_F^{\mbox{\tiny DM}}}}\hspace{-0.5cm}\frac{k^4dk}{(k^2+M^{*2}_\chi)^{1/2}}.
\label{pkindm}
\end{eqnarray}
About the nucleon kinetic energy contribution, we implement the modification in the momentum distribution function $n(k)$ by replacing the usual step function by the HMT one~\cite{cai,lucas}, in which 
\begin{eqnarray}
n_{n,p}(k) = \left\{ 
\begin{array}{ll}
\Delta_{n,p}, & 0<k<k_{F\,{n,p}}
\\ \\
\dfrac{C_{n,p}\,k_{F\,{n,p}}^4}{k^4}, & k_{F\,{n,p}}<k<\phi_{n,p} k_{F\,{n,p}}.
\end{array} 
\right.
\label{eqhtm}
\end{eqnarray}
Furthermore, $\Delta_{n,p}=1 - 3C_{n,p}(1-1/\phi_{n,p})$, where $C_p=C_0[1 - C_1(1-2y_p)]$, $C_n=C_0[1 + C_1(1-2y_p)]$, $\phi_p=\phi_0[1 - \phi_1(1-2y_p)]$ and $\phi_n=\phi_0[1 + \phi_1(1-2y_p)]$. Here we use $C_0=0.161$, $C_1=-0.25$, $\phi_0 = 2.38$ and $\phi_1=-0.56$~\cite{cai,lucas}. In Fig.~\ref{hmt} we plot this $n(k)$ distribution with HMT for some values of $\rho$, in units of saturation density $\rho_0$, for neutron matter ($y_p=0$).

The modification in the momentum integrals leads to 
\begin{eqnarray} 
\mathcal{E}_{\text {kin }}^{n,p} &=& \frac{\gamma \Delta_{n,p}}{2\pi^2} \int_0^{{k_{F\,{n,p}}}} 
k^2dk({k^{2}+M^{* 2}})^{1/2}
\nonumber\\
&+& \frac{\gamma C_{n,p}}{2\pi^2} \int_{k_{F\,{n,p}}}^{\phi_{n,p} {k_{F\,{n,p}}}} 
\frac{{k_F}_{n,p}^4}{k^2}\, dk({k^{2}+M^{* 2}})^{1/2},
\nonumber \\
P_{\text {kin }}^{n,p} &=&  
\frac{\gamma \Delta_{n,p}}{6\pi^2} \int_0^{k_{F\,{n,p}}}  
\frac{k^4dk}{\left({k^{2}+M^{*2}}\right)^{1/2}} 
\nonumber\\
&+& \frac{\gamma C_{n,p}}{6\pi^2} \int_{k_{F\,{n,p}}}^{\phi_{n,p} {k_{F\,{n,p}}}} 
 \frac{{k_F}_{n,p}^4dk}{\left({k^{2}+M^{*2}}\right)^{1/2}},
\end{eqnarray}
and
\begin{align}
&{\rho_s}_{n,p} = 
\frac{\gamma M^*\Delta_{n,p}}{2\pi^2} \int_0^{k_{F\,{n,p}}}  
\frac{k^2dk}{\left({k^{2}+M^{*2}}\right)^{1/2}} 
\nonumber\\
&+ \frac{\gamma M^*C_{n,p}}{2\pi^2} \int_{k_{F\,{n,p}}}^{\phi_{n,p} {k_{F\,{n,p}}}} 
\frac{{k_F}_{n,p}^4}{k^2}  \frac{dk}{\left({k^{2}+M^{*2}}\right)^{1/2}}.
\end{align}
This last quantity is the scalar density for protons and neutrons used in Eq.~(\ref{rhos}).

\section{Results} 
\label{res}

In this section, we present our results concerning the inclusion of dark matter in the thermodynamics of the relativistic mean-field hadronic model with short-range correlations. We assume the DM + hadron system described by the Lagrangian density from Eq.~(\ref{dlagtotal}) with the main EOS derived from it, taking into account the HMT model for the nucleon momentum distribution, as detailed in the previous section. 

\subsection{Hadronic parametrization with %dark matter 
DM content}
\label{part1}

At this point, it is needed to select a particular parametrization for the hadronic sector. We are interested in the influence of dark matter on models with SRC and its effects in the stellar matter. For this purpose, we choose the FSU2R parametrization proposed in Ref.~\cite{fsu2r} and updated in Ref.~\cite{fsu2r-new}. This new version, used in this section, was calibrated to reproduce properties of finite nuclei, as well as some nuclear matter constraints. In particular, among all neutron stars generated by this model, the heaviest one presents a mass of $2.05M_\odot$ ($M_\odot$ is the solar mass) with a corresponding radius of $11.6$~km. 
\begin{table}[!htb]
\centering
% \small
% \scriptsize
\caption{Coupling constants of the FSU2R parametrization with and without SRC included.}
\begin{tabular}{lrr}
\hline\noalign{\smallskip}
coupling &  FSU2R & FSU2R-SRC \\
\noalign{\smallskip}\hline\noalign{\smallskip}
$g_\sigma$  & $10.3718$ & $10.5174$ \\
$g_\omega$  & $13.5054$ & $12.3648$ \\
$g_\rho$    & $14.3675$ & $15.5988$ \\
$A/M_{\mbox{\tiny nuc}}$& $1.8365$ & $2.9133$ \\
$B$         & $-3.2403$ & $-32.4432$ \\
$\alpha_3'$ & $0.0900$ & $0.0093$ \\
\noalign{\smallskip}\hline
\label{tabconst}
\end{tabular}
\end{table}

In Tab.~\ref{tabconst}, we provide the coupling constants of the parametrization with SRC (\mbox{FSU2R-SRC}) and without it~(FSU2R). Moreover, we also take $C=0.004$, $M_{\mbox{\tiny nuc}}=939$~MeV, $m_\sigma=497.479$~MeV, $m_\omega=782.5$~MeV, $m_\rho=763$~MeV for both versions.
We emphasize that the values of the coupling constants presented in Table~\ref{tabconst} were found by imposing the same bulk parameters for both approaches (model with and without SRC), namely, $\rho_0=0.15$~fm$^{-3}$, $B_0=-16.27$~MeV (binding energy), $m^*=M^*_0/M_{\mbox{\tiny nuc}}=0.593$ ($M_0^*$: effective nucleon mass at $\rho=\rho_0$), $K_0=237.7$~MeV (incompressibility at $\rho=\rho_0$), $J=30.7$~MeV (symmetry energy at $\rho=\rho_0$), and $L_0=46.9$~MeV (symmetry energy slope at $\rho=\rho_0$). These are exactly the bulk parameters of the ``reference model'' FSU2R~\cite{fsu2r-new}.

\begin{figure}[!htb] 
\centering
\includegraphics[scale=0.33]{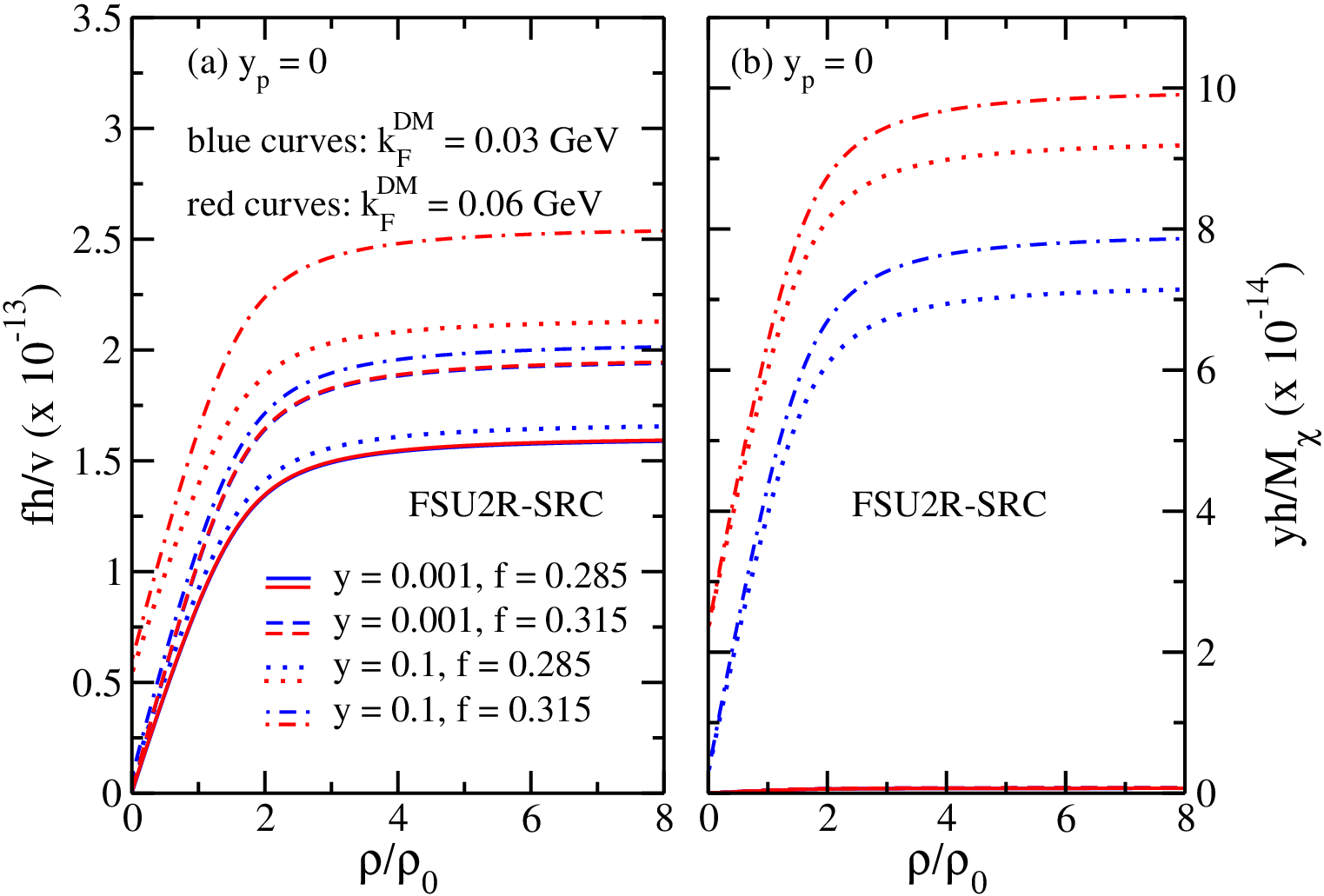}
\includegraphics[scale=0.33]{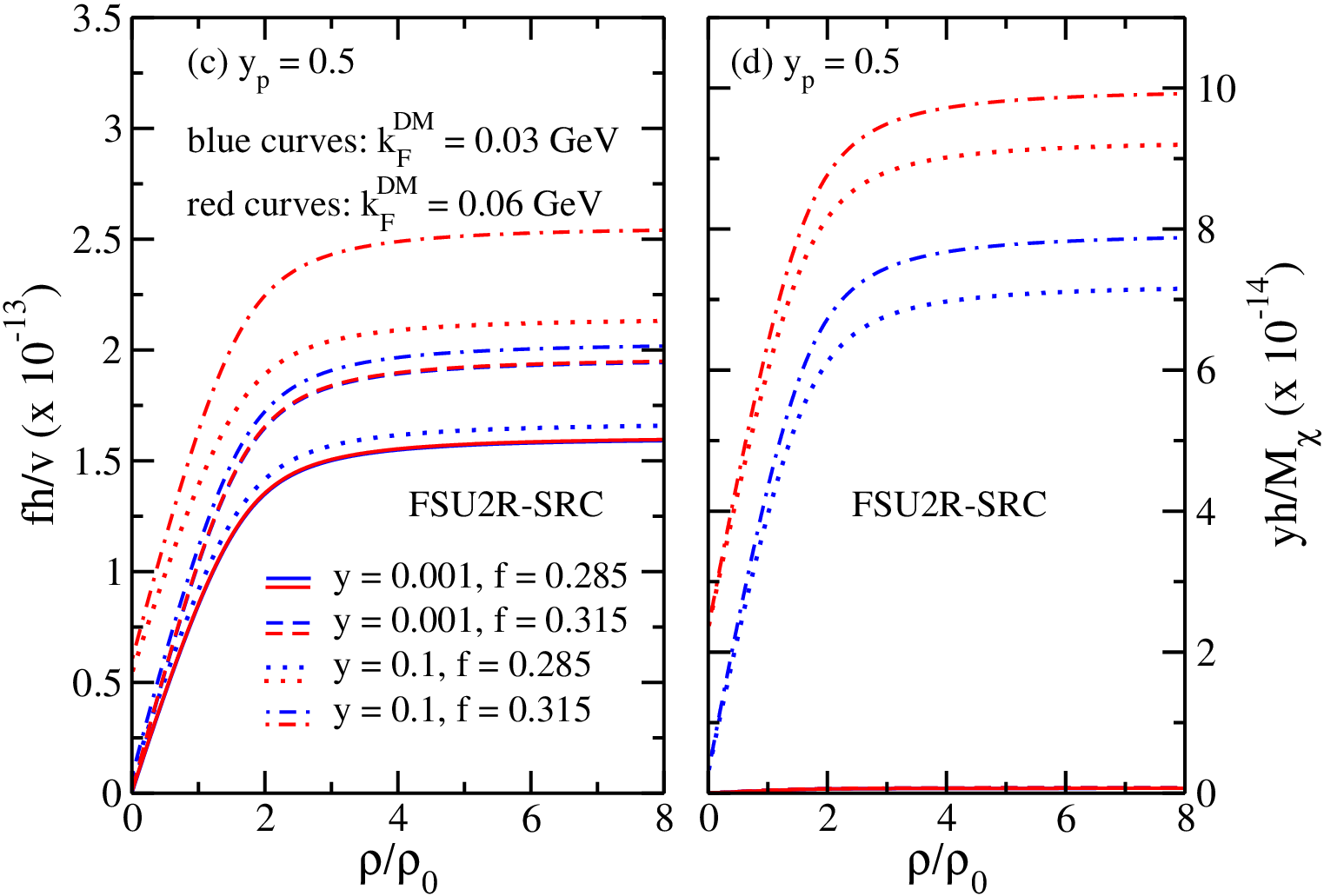}
\caption{$fh/v$ and $yh/M_\chi$ as function of $\rho/\rho_0$ for neutron matter ($y_p=0$: top panels) and symmetric matter ($y_p=0.5$: bottom panels)  with different Fermi momentum of the dark particle.}
\label{higgs1}
\end{figure}

We start by treating the whole system composed of hadrons and heavy dark particles, namely, the lightest neutralino, with mass $M_\chi=200$~GeV. The mediator between the dark sector and the hadronic one is the Higgs boson, having a mass of $m_h=125$~GeV. This approach was used in Refs.~\cite{rmfdm2,rmfdm3,rmfdm4,rmfdm8,rmfdm9}, for instance, with hadronic models without SRC included. Regarding the couplings of the dark matter sector, we base our study in values of $y$ and $f$ inside the ranges of $0.001\leqslant y\leqslant 0.1$~\cite{rmfdm2}, and $f=0.3\pm0.015$~\cite{cline}. In order to investigate the effect of these parameters in the hadronic-DM matter model, we present in Fig.~\ref{higgs1} the magnitude of the quantities $fh/v$ and $yh/M_\chi$, plotted as a function of the density, for different values of the proton fraction and $k_F^{\mbox{\tiny DM}}$ using a combination of the boundary values of $y$ and $f$. 
From the results presented in this figure, one can conclude that~$M^*$ and~$M^*_\chi$ are practically not affected by the Higgs field, since $fh/v$ and $yh/M_\chi$ are around $10^{-13}$ and $10^{-14}$, respectively. Therefore, the effective nucleon mass keeps the usual form obtained in the RMF models, i.e., depending only on the scalar field $\sigma$, $M^*\simeq M_{\mbox{\tiny nuc}}-g_\sigma\sigma$, see Eq.~(\ref{mnuc}). Furthermore, it is also clear that $M^*_\chi$ can be safely given by its vacuum value, namely, $M^*_\chi\simeq M_\chi$, according to Eq.~(\ref{mneu}).

\begin{figure}[!htb] 
\centering
\includegraphics[scale=0.33]{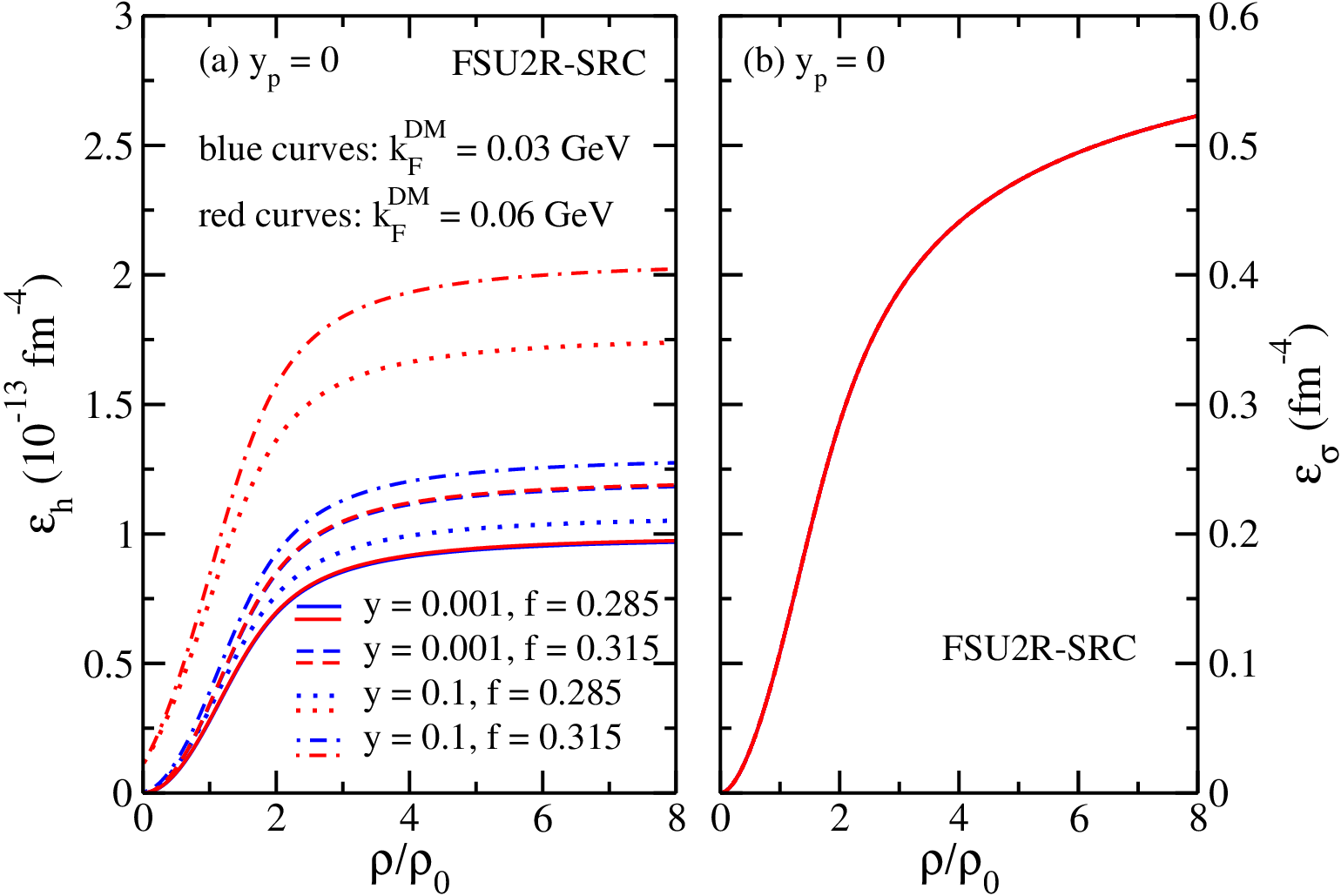}
\includegraphics[scale=0.33]{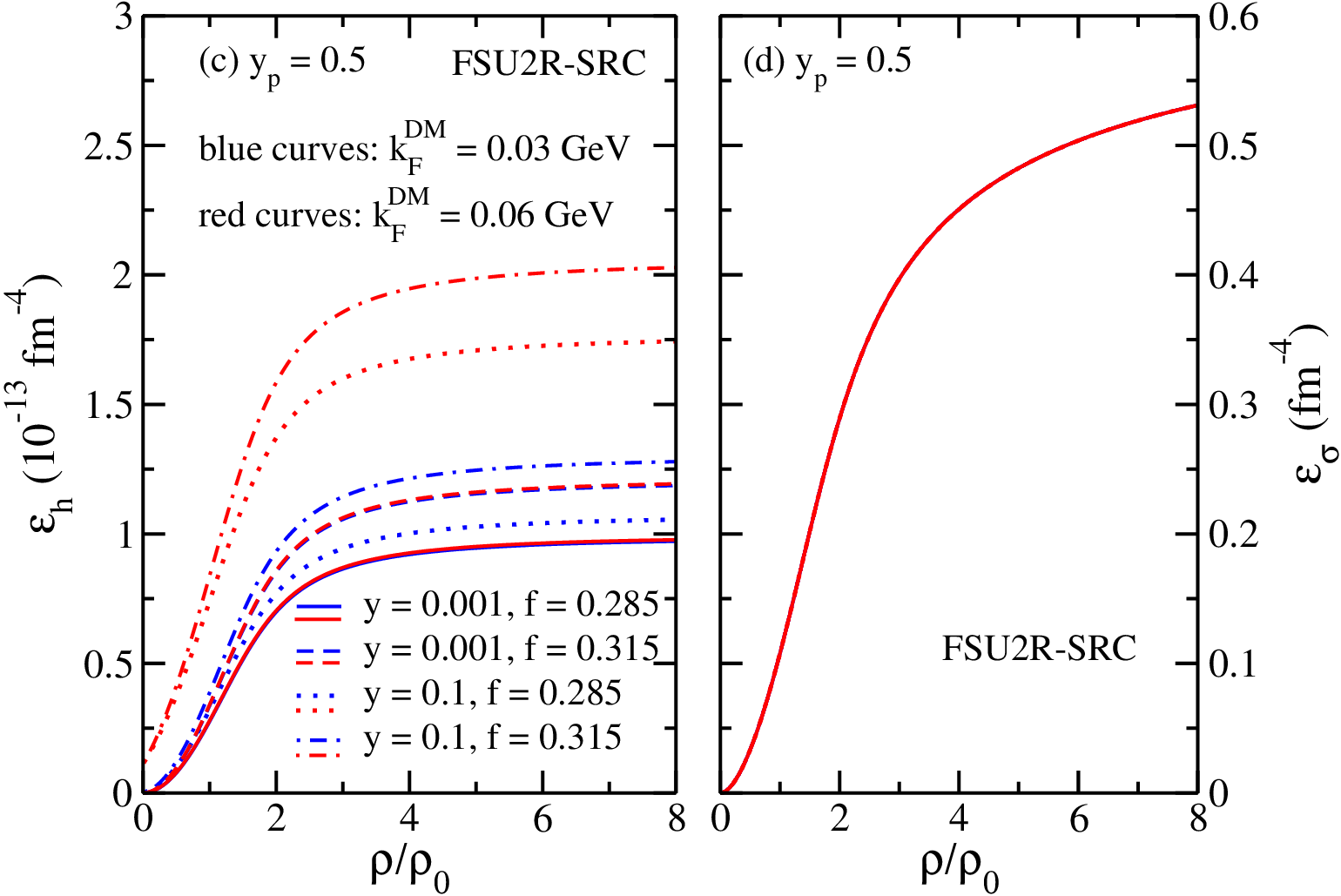}
\caption{$\mathcal{E}_h$ and $\mathcal{E}_\sigma$ as function of $\rho/\rho_0$ for neutron matter ($y_p=0$: top panels) and symmetric matter ($y_p=0.5$: bottom panels) with different Fermi momentum of the dark particle.}
\label{higgs2}
\end{figure}

We also investigate the thermodynamical equations of state given in Eqs.~(\ref{eden}) and (\ref{press}), specifically regarding the contribution of the terms involving the Higgs field. By defining the quantity $\mathcal{E}_h=m_h^2h^2/2$, we verify in Fig.~\ref{higgs2} the magnitude of this term, as a function of $\rho/\rho_0$, in comparison with $\mathcal{E}_\sigma=m_\sigma^2\sigma^2/2$, for instance. 
It is clear that $\mathcal{E}_h\ll\mathcal{E}_\sigma$, since $\mathcal{E}_h\simeq 10^{-13}$~fm$^{-4}$. Therefore, it is justified to discard $\mathcal{E}_h$ in Eqs.~(\ref{eden}) and (\ref{press}) in comparison with the other terms. This is an important feature to the calculation of the chemical potentials of protons and neutrons. Since $\mu_{p,n}=\partial\mathcal{E}/\partial\rho_{p,n}$, in principle the term $\mathcal{E}_h$ would contribute to the final expressions. Furthermore, the derivatives of protons and neutrons would also act on $\mathcal{E}_{\mbox{\tiny kin}}^{\mbox{\tiny DM}}$ and $P_{\mbox{\tiny kin}}^{\mbox{\tiny DM}}$ due to $M^*_\chi$. However, since it is verified that we can use $M^*_\chi\simeq M_\chi$, $M^*\simeq M_{\mbox{\tiny nuc}}-g_\sigma\sigma$, and also neglect $\mathcal{E}_h$ in Eq.~(\ref{eden}), we recognize that the nucleon derivatives only affects the hadronic part of the energy density. The consequence is that $\mu_{p,n}$ can be given by the same expressions related exclusively to the hadronic sector. In the case of the model with SRC included, they are
\begin{eqnarray} 
&\mu_{p,n}& = 3 C_{p,n} \left[ \mu^{p,n}_{\mathrm{kin}}
- \frac{\left({\phi_{p,n}^2 {k^2_F}_{p,n} + M^{*2}}\right)^{1/2}}{\phi_{p,n}} \right]
\nonumber\\
&+& {4}C_{p,n} {k_F}_{p,n} \ln\left[\frac{\phi_{p,n} {k_F}_{p,n} + 
\left(\phi_{p,n}^2{k_F^2}_{p,n}+M^{*2}\right)^{1/2} }{ {k_F}_{p,n} + \left(  {k^2_F}_{p,n} + M^{*2}\right)^{1/2}}\right] 
\nonumber\\
&+& \Delta_{p,n}\mu^{p,n}_{\mathrm{kin}} + g_{\omega} \omega_{0} \pm \frac{g_\rho}{2}\bar{\rho}_{0_{(3)}},
\end{eqnarray} 
with $\mu^{p,n}_{\mathrm{kin}}=({k^2_F}_{p,n}+M^{*2})^{1/2}$.

From the discussion of the results presented in Figs.~\ref{higgs1} and~\ref{higgs2}, we verify that $y$ and $f$ do not play a significant role in the thermodynamics of the DM-hadron system. It means that the DM contribution can be described only by the dark fermion kinetic terms. In other words, the particular choice of the pair $y$ and $f$ does not affect the results we will present in what follows. Therefore, we take $y=0.01$ since this value, combined to $M_\chi=200$~GeV, produces a spin-independent scattering cross-section around $10^{-47}$~cm$^2$~\cite{rmfdm4}, in agreement with experimental boundaries obtained by PandaX-II~\cite{pandaxII}, LUX~\cite{lux}, and DarkSide~\cite{darkside} collaborations regarding direct detection experiments. We also use $f=0.3$, namely, the central value obtained in Ref.~\cite{cline}.

\begin{figure}[!htb] 
\centering
\includegraphics[scale=0.33]{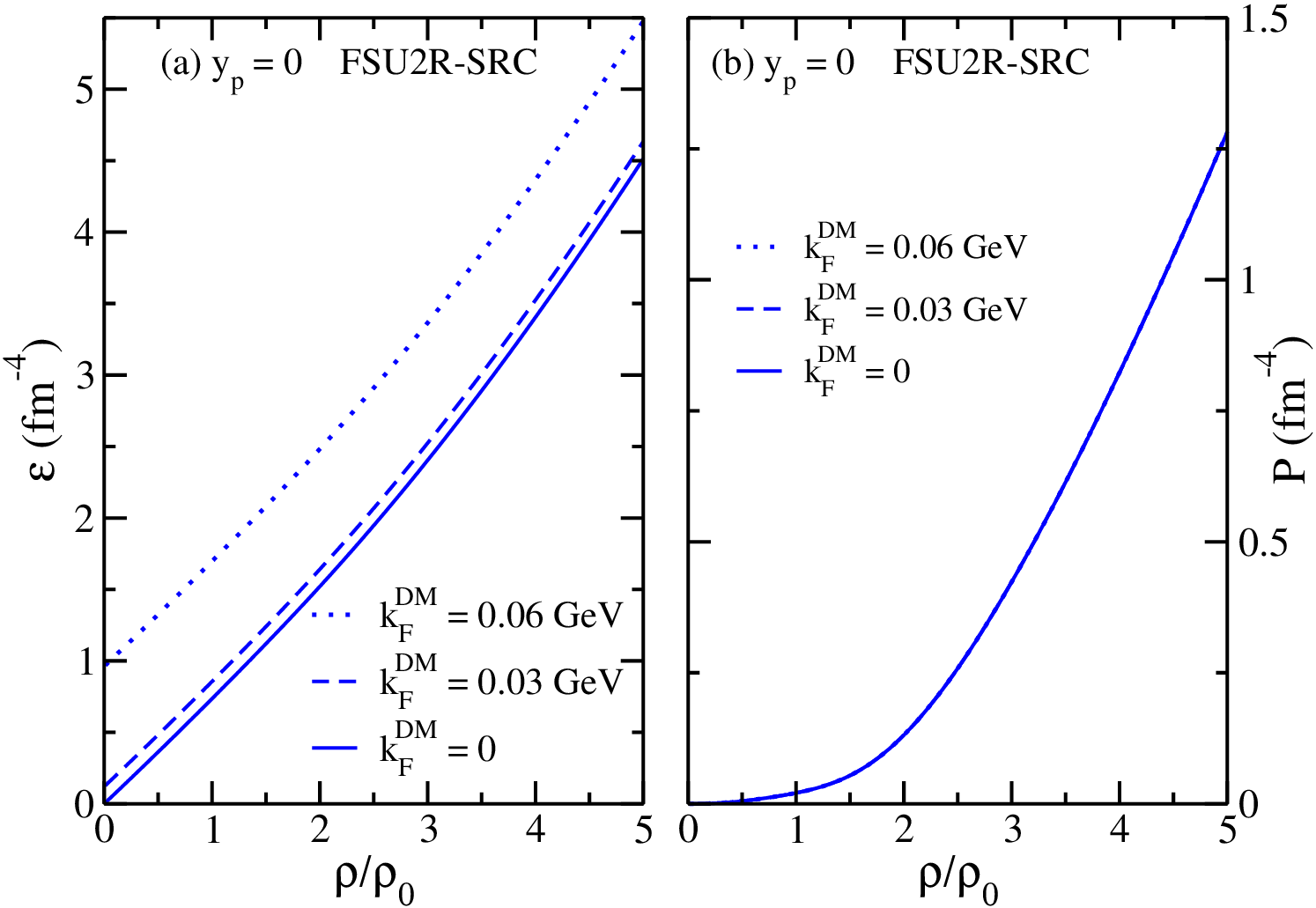}
\caption{(a) $\mathcal{E}$ and (b) $P$ as functions of $\rho/\rho_0$ for neutron matter ($y_p=0$). Results for different Fermi momenta of the dark particle.}
\label{eoskfdm}
\end{figure}

For the sake of completeness, we show in Fig.~\ref{eoskfdm} the density dependence of Eqs.~(\ref{eden}) and~(\ref{press}) for neutron matter ($y_p=0$). 
In the left panel of the figure, we observe that the energy density is quite sensitive to the DM content of the system even at low densities. For the pressure, shown in the right panel of the figure, there is no significant contribution. This follows from the small values of the DM Fermi momentum and the large neutralino mass, as already discussed. Therefore, since $k_F^{\mbox{\tiny DM}}$ is fixed, one has $\mathcal{E}_{\mbox{\tiny kin}}^{\mbox{\tiny DM}}$ constant and $P_{\mbox{\tiny kin}}^{\mbox{\tiny DM}}\simeq 0$ (see Eqs.~\ref{ekindm} and~\ref{pkindm}). The result of this combination is the shift in the energy density curves with no variation in the respective pressure curves. This result is compatible with the findings presented in Ref.~\cite{rmfdm6}, for instance.

In order to analyze the effect of dark matter coupled to the RMF-SRC model in the astrophysical context, more specifically, the description of neutron stars, we consider stellar matter under charge neutrality and $\beta$-equilibrium, i.e., a system in which the weak process and its inverse 
reaction, namely, $n\rightarrow p + e^- + \bar{\nu}_e$ and $p+ e^-\rightarrow n + \nu_e$, occur simultaneously. Besides massless electrons, we also consider muons, that appear when the electron chemical potential $\mu_e$ exceeds the muon mass ($m_\mu=105.7$~MeV). In this case, the following conditions holds: $\rho_p-\rho_e=\rho_\mu$ and $\mu_n-\mu_p=\mu_e$, with $\mu_\mu=\mu_e$. $\rho_e$ is the electron density with $\mu_e$ and $\rho_e$ related to each other through $\rho_e=\mu_e^3/(3\pi^2)$. The muon density is $\rho_\mu=[(\mu_\mu^2 - m_\mu^2)^{3/2}]/(3\pi^2)$. By considering these two leptons in system, the total energy density and pressure read
\begin{eqnarray}
\epsilon = \mathcal{E} + \frac{\mu_e^4}{4\pi^2} 
+ \frac{1}{\pi^2}\int_0^{\sqrt{\mu_\mu^2-m^2_\mu}}\hspace{-0.6cm}dk\,k^2(k^2+m_\mu^2)^{1/2},
\label{totaled}
\end{eqnarray}
and
\begin{align} 
p = P + \frac{\mu_e^4}{12\pi^2} +\frac{1}{3\pi^2}\int_0^{\sqrt{\mu_\mu^2-m^2_\mu}}\hspace{-0.5cm}\frac{dk\,k^4}{(k^2+m_\mu^2)^{1/2}},
\label{totalp}
\end{align}
with $\mathcal{E}$ and $P$ obtained in Eqs.~(\ref{eden}) and (\ref{press}). These equations of state are used as input to solve the Tolman-Oppenheimer-Volkoff (TOV) equations~\cite{tov39,tov39a,glen} given by~($G=c=1$) 
\begin{align}
\frac{dp(r)}{dr}&=-\frac{[\epsilon(r) + p(r)][m(r) + 4\pi r^3p(r)]}{r^2[1-2m(r)/r]},
\label{tov1}
\\
\frac{dm(r)}{dr}&=4\pi r^2\epsilon(r),
\label{tov2}
\end{align}
whose solution is constrained to $p(0)=p_c$ (central pressure) and $m(0) = 0$. At the star surface, one has $p(R) = 0$ and $m(R)\equiv M$, with $R$ defining its radius. In regard to the neutron star crust, we model this specific region by splitting it in two parts. The first one, the outer crust, is described by the Baym-Pethick-Sutherland (BPS) equation of state~\cite{bps} in the 
density region of $6.3\times10^{-12}\,\mbox{fm}^{-3} \leqslant\rho\leqslant 
2.5\times10^{-4}\,\mbox{fm}^{-3}$~\cite{poly2,malik19}. The second part is the inner crust. For this specific region we use the polytropic form given by $p(\epsilon)=A+B\epsilon^{4/3}$~\cite{poly2,poly1,gogny2} in a range of 
$2.5\times10^{-4}\,\mbox{fm}^{-3} \leqslant\rho\leqslant \rho_t$, where $\rho_t$ (transition density) is the density associated to the core-crust transition, found here through the thermodynamical method~\cite{gogny1,cc2,kubis04,gonzalez19}. 

\begin{figure}[!htb] 
\centering
\includegraphics[scale=0.33]{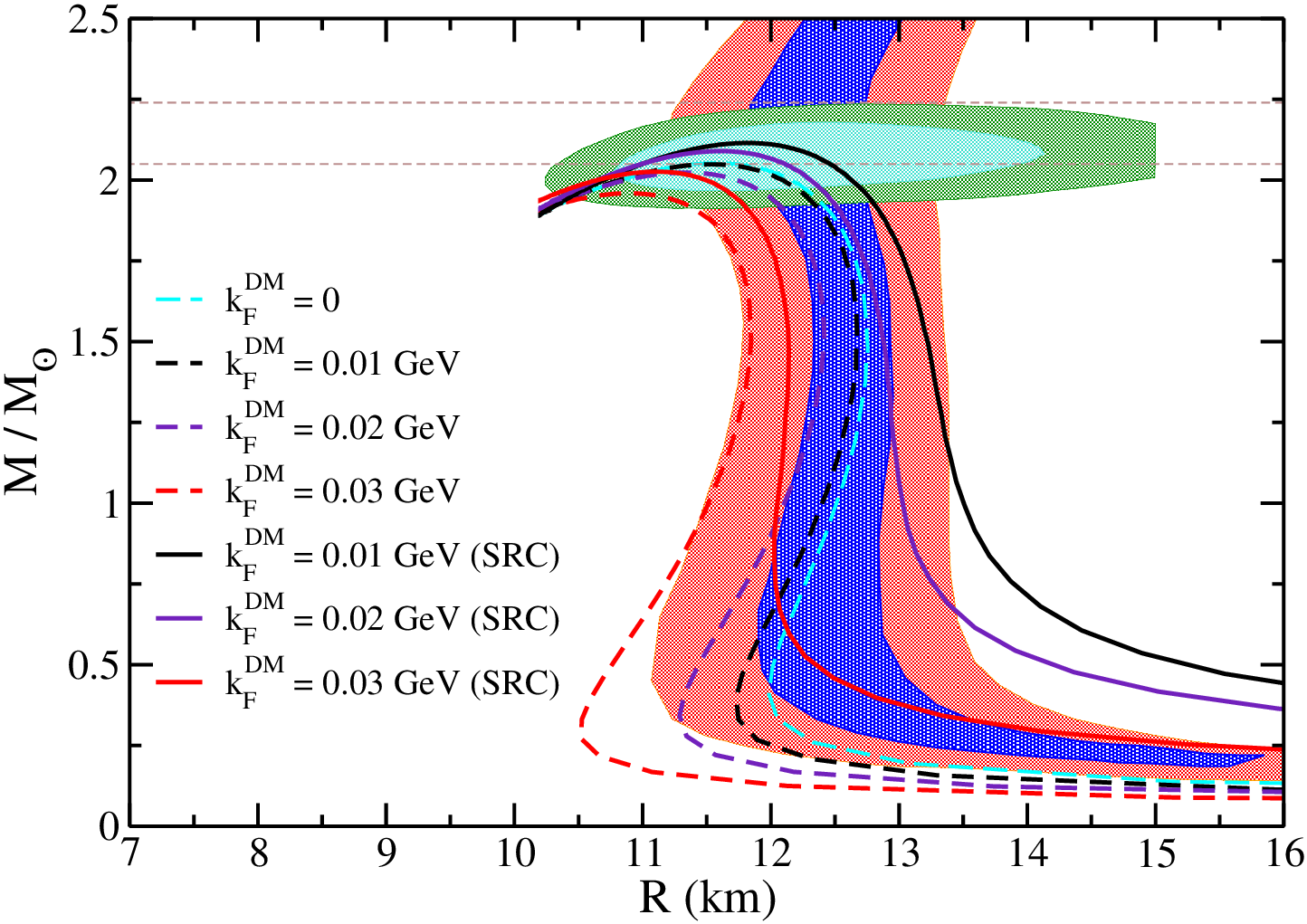}
\caption{Neutron star mass over $M_\odot$ (solar mass) as a function of its radius obtained from 
the DM-RMF model, with and without SRC included, for different values of $k_F^{\mbox{\tiny DM}}$. 
The bands extracted from Refs.~\cite{arx79,arx80} are the regions constructed in recent analyses of 
the data coming from NICER mission, LIGO and Virgo collaborations, and radio observations. The 
region delimited by the horizontal dashed lines was extracted from Ref.~\cite{cromartie}.} 
\label{mr1}
\end{figure}

In Fig.~\ref{mr1} we display the mass-radius profiles of the DM  model coupled to the RMF one with and without SRC included. 
In this figure, we compare our results with the bands determined from the result of multi-messenger 
analyses performed by in Refs.~\cite{arx79,arx80} regarding the PSR J0740+6620 pulsar. Blue (25th to 
75th percentile range) and red (5th to 95th) bands take into account symmetry energy measurements, 
tidal deformability upper limits, and mass-radius posteriors on the PSR J0030+0451 and PSR 
J0740+6620 pulsars, based on the NICER and X-ray Multi-Mirror (XMM-Newton) X-ray 
observations~\cite{arx79}. The green bands are the $68\%$ (internal) and $95\%$ credible regions for 
mass and radius inferred from NICER and XMM-Newton European Photon Imaging Camera data~\cite{arx80}. 
In summary, besides the NICER data, these analyses also include gravitational wave data obtained 
from LIGO and Virgo collaborations and maximum mass measurements from radio observations. Our 
findings are also compared with the range given by Ref.~\cite{cromartie} at $68.3\%$ credible level 
and also related to the J0740+6620, namely, $M=2.14^{+0.10}_{-0.09}M_{\odot}$ (horizontal dashed 
lines). 

\begin{figure}[!htb] 
\centering
\includegraphics[scale=0.33]{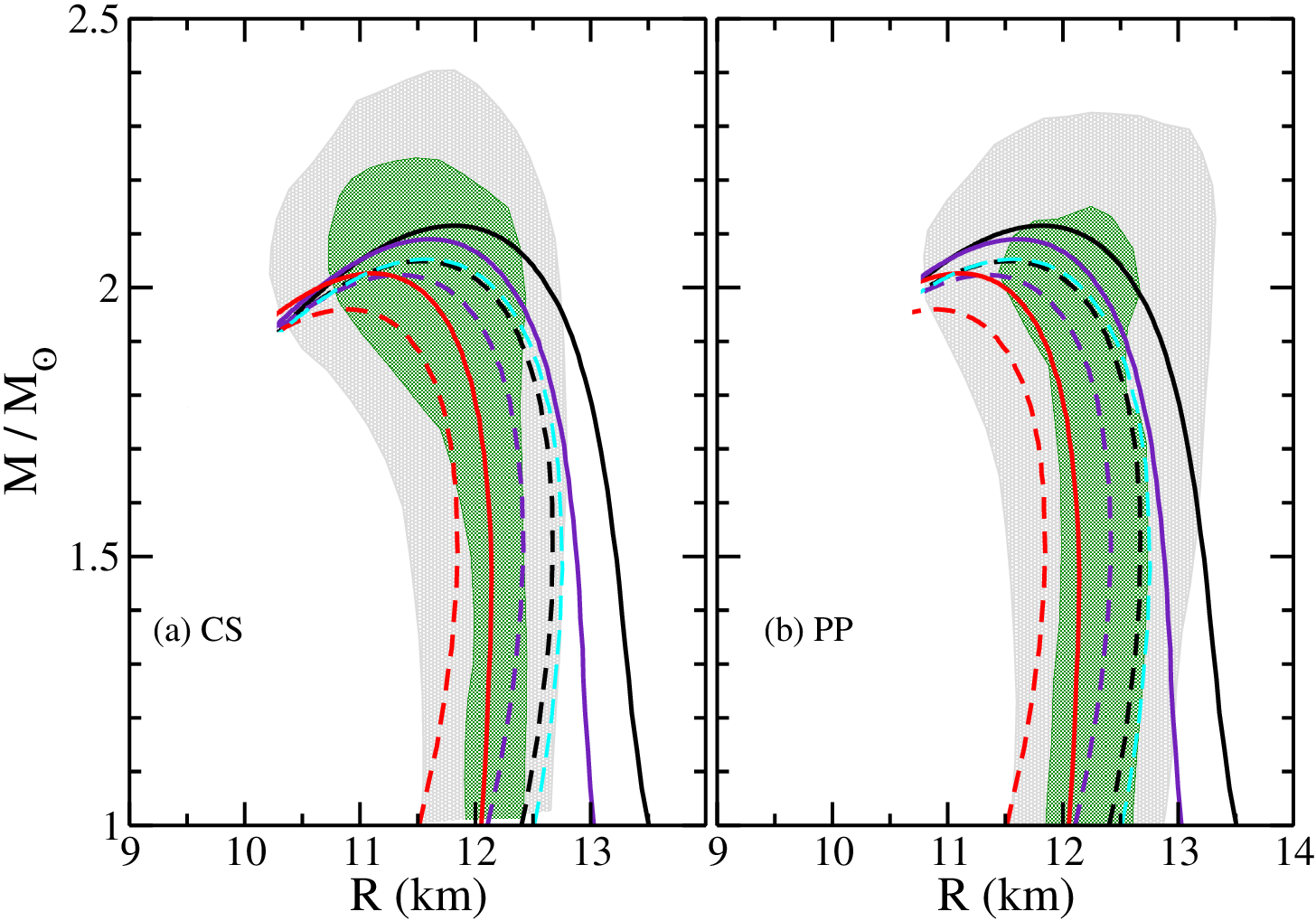}
\caption{Curves: the same as in Fig.~\ref{mr1}. The two possible different regions for the star mass and radius are extracted from figure~5 of Ref.~\cite{arx81}. These regions are obtained by analyzing the data with a model based on the speed of sound in a neutron star (panel a), and  with the piecewise-polytropic model (panel b).} 
\label{mr2}
\end{figure}

In Fig.~\ref{mr2} we compare the same curves shown in Fig.~\ref{mr1} with other data extracted from figure~5 of Ref.~\cite{arx81}. More specifically, it is shown the $68\%$ (internal band) and $95\%$ credible regions from the jointly analysis of mass-radius estimates from PSR J0740+6620 and PSR J0030+0451 pulsars, mass-tidal deformability estimates from GW170817  and GW190425 events, and AT2017gfo kilonova data~\cite{arx81}. These regions were obtained from two different approaches employed in the analysis presented by Ref.~\cite{arx81}, namely, a model based on the speed of sound in a neutron star, (CS, Fig.~\ref{mr2}{\color{blue}a}) and the piecewise-polytropic model (PP, Fig.~\ref{mr2}{\color{blue}b}).

\begin{figure}[!htb] 
\centering
\includegraphics[scale=0.33]{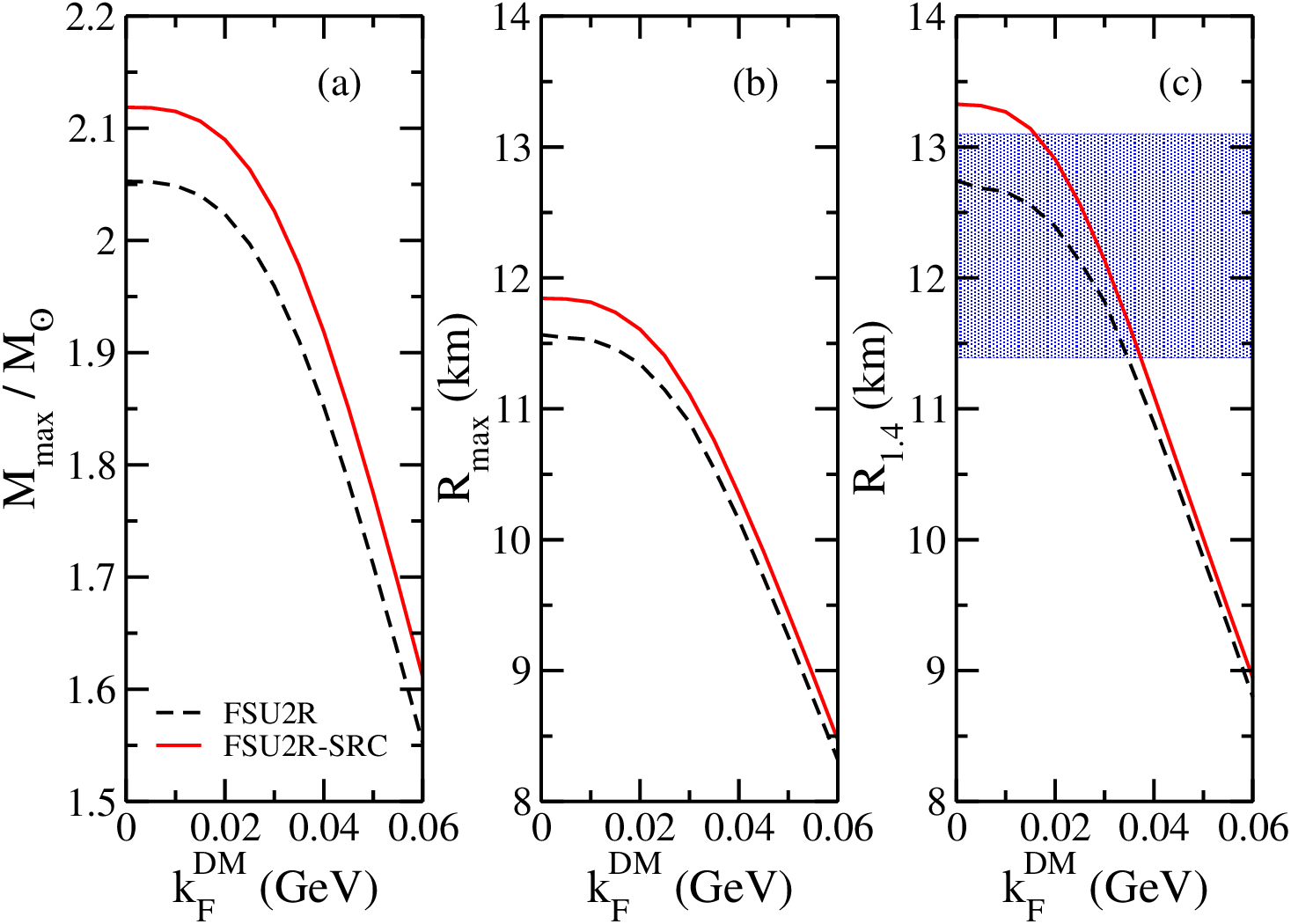}
\caption{Maximum neutron star (a) mass, (b) radius, and radius of the $M=1.4M_\odot$ neutron star. All of them were plotted as a function of DM Fermi momentum. Blue region: data extracted from Refs.~\cite{arx79,arx81}.} 
\label{kfdm}
\end{figure}

Some interesting features can be observed from Figs.~\ref{mr1} and \ref{mr2}. The first one is concerning the possibility of a description of neutron stars with some content of dark matter. Our findings indicate that there is a certain range for $k_F^{\mbox{\tiny DM}}$ that produces mass-radius diagrams compatible with the recent observations, and simultaneously, with the range of $M=2.14^{+0.10}_{-0.09} M_{\odot}$ from Ref.~\cite{cromartie}. Another important result is the effect of SRC in the DM matter model coupled to the hadronic one. It is known that increasing the dark matter content in the system, i.e.,  by growing  $k_F^{\mbox{\tiny DM}}$, softens the equation of state~\cite{rmfdm2,rmfdm3,rmfdm6,rmfdm8,rmfdm9}. As a direct consequence, there is, in principle, a conflict between the existence of DM in neutron stars and the possibility that these compact astrophysical objects attain higher mass values. Notice that the current observational data points out values greater than $2M_\odot$ for this quantity. However, it is also known that the inclusion of SRC in hadronic models produces exactly the opposite effect~\cite{cai,lucas}, i.e, models with SRC become stiffer and because of that, they can describe more massive stars in comparison with those without this effect (for models with the same bulk properties). Therefore, SRC can balance the decreasing of the neutron star mass due to the inclusion of DM in the system. This is exactly what we observe in the last two figures. The comparison of the DM-RMF model (dashed curves) with DM-RMF-SRC one (full lines) shows a shift of the maximum neutron star masses in the direction of higher values. The same kind of increase is also observed in the neutron star radius. Such an effect can be better visualized in Fig.~\ref{kfdm}.

\begin{figure}[!htb] 
\centering
\includegraphics[scale=0.3]{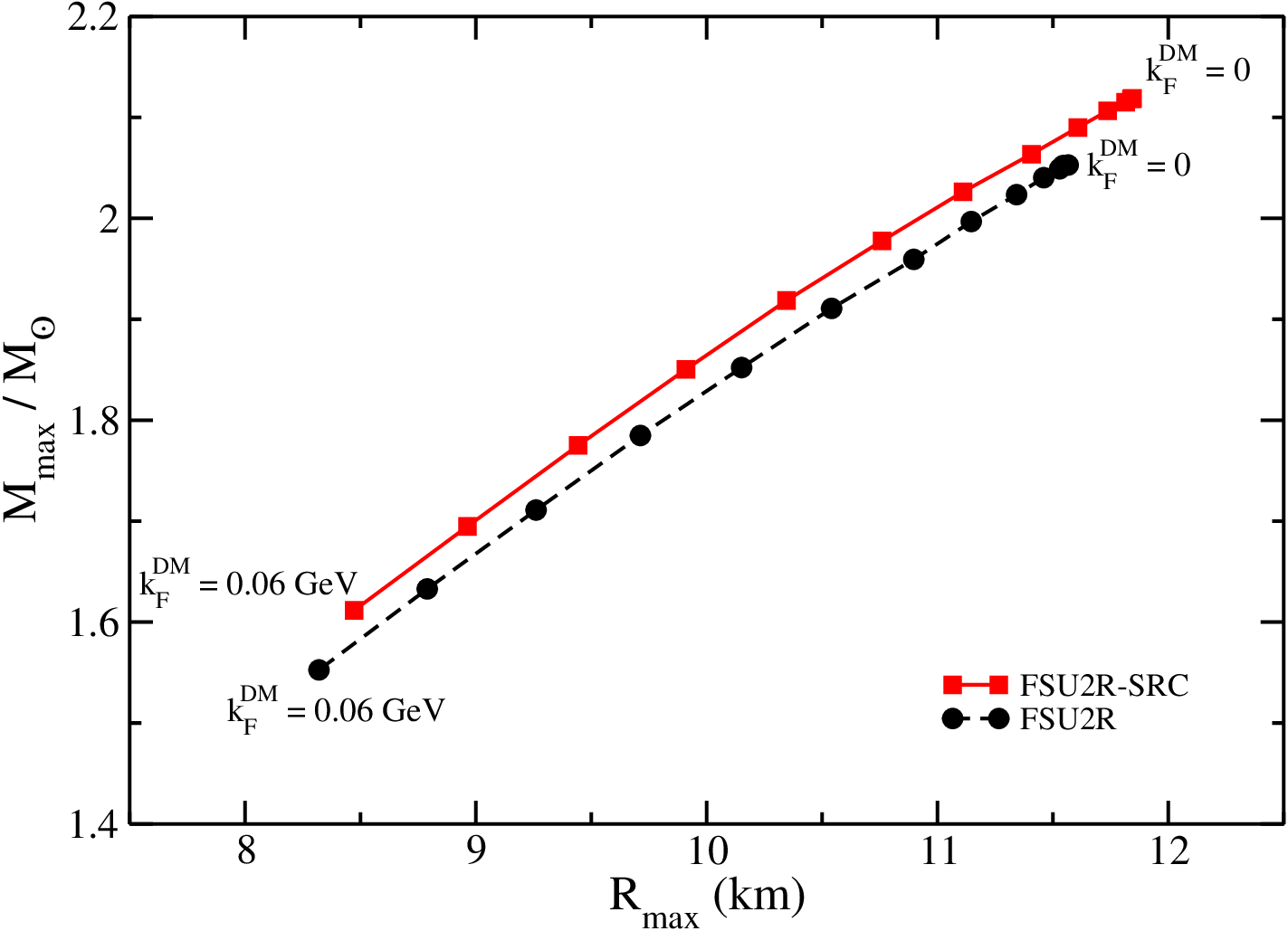}
\caption{$M_{\mbox{\tiny max}}$ versus $R_{\mbox{\tiny max}}$ for DM matter coupled to RMF model with and without SRC included.} 
\label{mrmax}
\end{figure}

From Fig.~\ref{kfdm}, one can clearly notice the effect of the SRC in the DM-RMF model concerning the values of the neutron star maximum mass ($M_{\mbox{\tiny max}}$), its corresponding radius ($R_{\mbox{\tiny max}}$), and the radius of the $M=1.4M_\odot$ neutron star ($R_{1.4}$). In particular, in panel (c) we also display the band region predicted by the combination of the recent data given in Refs.~\cite{arx79,arx81}. In regard to the findings shown in panels (a) and (b), we also present them from another perspective in Fig.~\ref{mrmax}.
From this figure, besides a clear correlation between $M_{\mbox{\tiny max}}$ and $R_{\mbox{\tiny max}}$, we also verify that the %entire 
whole curve produced by the model with SRC included is moved up and right in comparison to the one in which this phenomenology is absent. In summary, the decrease of $M_{\mbox{\tiny max}}$ and $R_{\mbox{\tiny max}}$ as $k_F^{\mbox{\tiny DM}}$ grows is balanced, to some extend, by SRC.

%\subsection{Effect of the bulk parameters variation in the system with dark matter content}
\subsection{Bulk parameters  uncertainties}
\label{part2}

\begin{figure}[!htb] 
\centering
\includegraphics[scale=0.364]{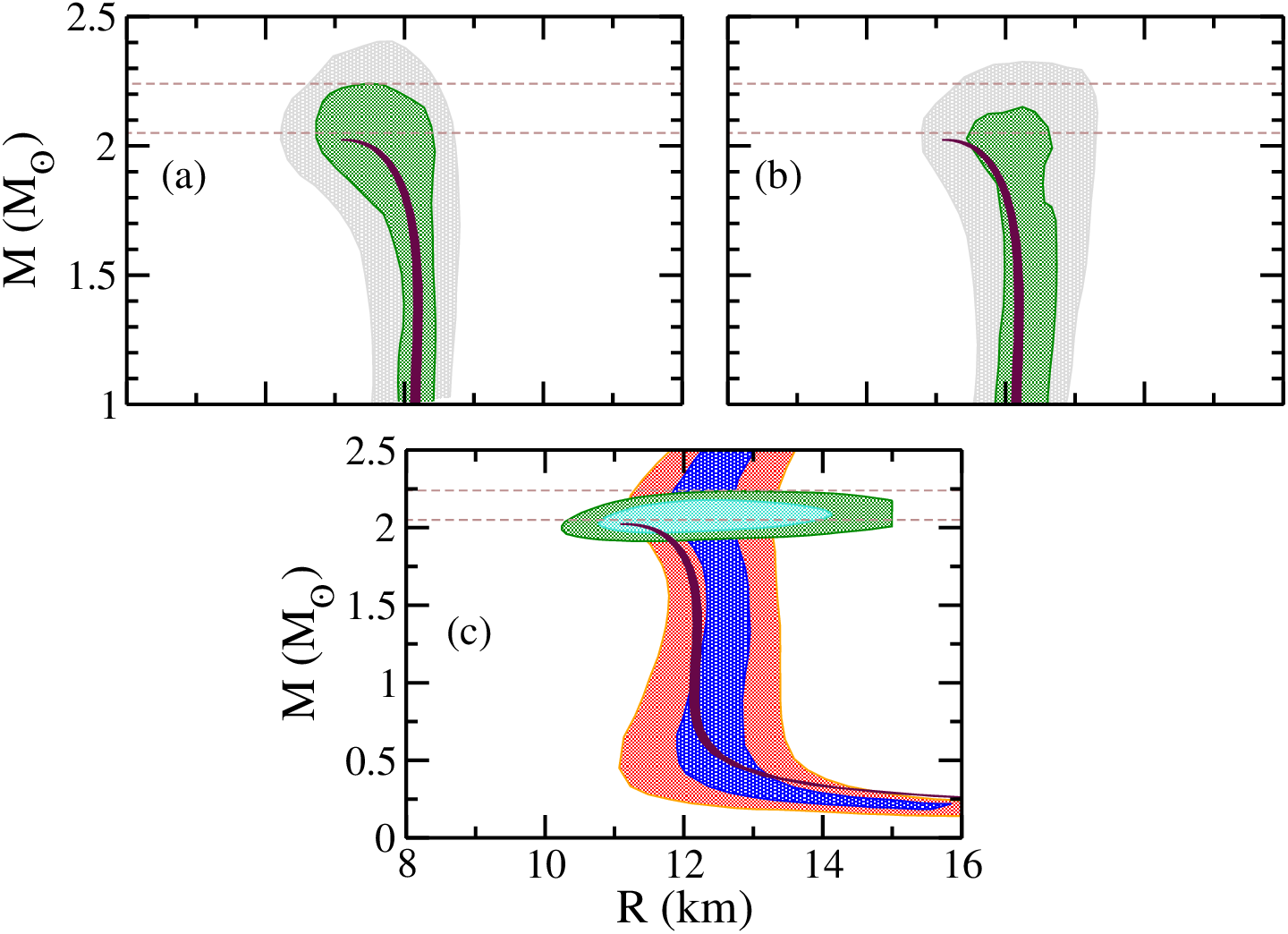}
\caption{Mass-radius diagrams. Brown region: band generated by the variation $220\,\mbox{MeV}\leqslant K_0 \leqslant 260~\mbox{MeV}$ in the \mbox{RMF-SRC} model with DM content ($k_F^{\mbox{\tiny DM}}=0.03$~GeV). Other regions: the same as in Figs.~\ref{mr1} and~\ref{mr2}. Horizontal lines: data taken from Ref.~\cite{cromartie}.}
\label{bandk0}
\end{figure}

Another study performed here is the analysis of the effect on stellar matter quantities caused by the uncertainties of some bulk parameters of the model composed by nucleons, leptons, and dark matter. In order to do that, we take the bulk parameter values of our reference model (\mbox{FSU2R}) as the starting point, see Table~\ref{tabbulk}. 
\begin{table}[!htb]
\centering
% \small
% \scriptsize
\caption{Some bulk parameters of the reference model. Here $\tilde{J}\equiv\mathcal{S}(2\rho_0/3)$.}
\begin{tabular}{cccccc}
\hline\noalign{\smallskip}
$\rho_0$ (fm$^{-3}$) & $B_0$ (MeV) & $K_0$ (MeV) & $m^*$ & $\tilde{J}$ (MeV) & $L_0$ (MeV)\\
\noalign{\smallskip}\hline\noalign{\smallskip}
$0.15$ & $-16.27$ & $237.7$ & $0.593$ & $25.68$ & $46.9$ \\
\noalign{\smallskip}\hline
\label{tabbulk}
\end{tabular}
\end{table}
Specifically, we use $\rho_0=0.15$~fm$^{-3}$ and $B_0=-16.27$~MeV as fixed in all calculations, since these numbers are well established in the literature close to these values. This does not apply to the other quantities such as the incompressibility, nucleon effective mass, symmetry energy, and its slope. For the first, we take the range of $220\,\mbox{MeV}\leqslant K_0 \leqslant 260~\mbox{MeV}$~\cite{garg,apj19}. The band obtained from this variation range is depicted in Fig.~\ref{bandk0}.
In order to construct it, we generate different parametrizations with SRC included from the 
independent variation only of $K_0$, with the other quantities fixed in the values presented in 
Table~\ref{tabbulk}. We also use $k_F^{\mbox{\tiny DM}}=0.03$~GeV, which gives $\rho^{\mbox{\tiny 
DM}}/\rho_0\simeq 10^{-3}$. This value also ensures that the ratio between the DM mass and the total 
neutron star mass is around $1/6$, according to Refs.~\cite{rmfdm2,rmfdm3}. From the figure, we 
verify that the variation in $K_0$ produces a very narrow band in the mass-radius diagram. It is 
also worth noticing that despite reaching the regions determined from NICER, LIGO and Virgo, and 
data from radio observations, the boundaries of $M=2.14^{+0.10}_{-0.09}M_{\odot}$~\cite{cromartie} 
are not reached.

\begin{figure}[!htb] 
\centering
\includegraphics[scale=0.364]{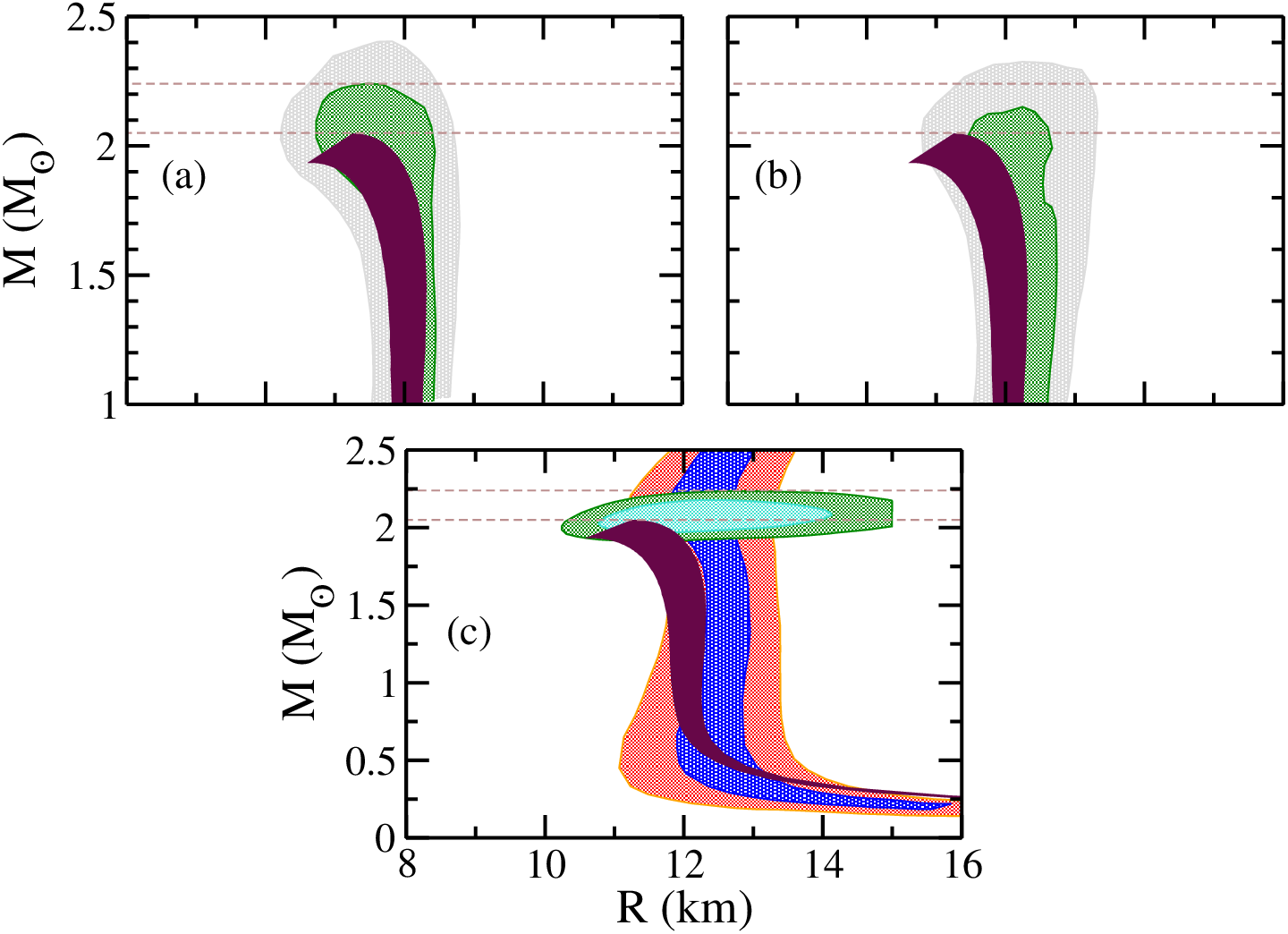}
\caption{Mass-radius diagrams. Brown region: band generated by the variation $0.58\leqslant m^* \leqslant 0.64$ in the \mbox{RMF-SRC} model with DM content ($k_F^{\mbox{\tiny DM}}=0.03$~GeV). Other regions: the same as in Figs.~\ref{mr1} and~\ref{mr2}. Horizontal lines: data taken from Ref.~\cite{cromartie}.}
\label{bandm}
\end{figure}

The same analysis was made by taking the variation of another isoscalar quantity, namely, the effective nucleon mass ratio, $m^*=M^*_0/M_{\mbox{\tiny nuc}}$. The interval
we assume for this quantity is given by $0.58\leqslant m^* \leqslant 0.64$. It is 
claimed to produce spin-orbit splittings in agreement with well-established experimental values for $^{16}\rm O$, $^{40}\rm Ca$, and $^{208}\rm Pb$ nuclei, according to the findings of Ref.~\cite{furns}. The band found from this specific variation, with the remaining bulk parameters kept fixed at the values of Table~\ref{tabbulk}, is shown in Fig.~\ref{bandm}. 
The results show a wider band in comparison with the previous one. However, the limits of $M=2.14^{+0.10}_{-0.09}M_{\odot}$~\cite{cromartie} are still not reached.

\begin{figure}[!htb] 
\centering
\includegraphics[scale=0.31]{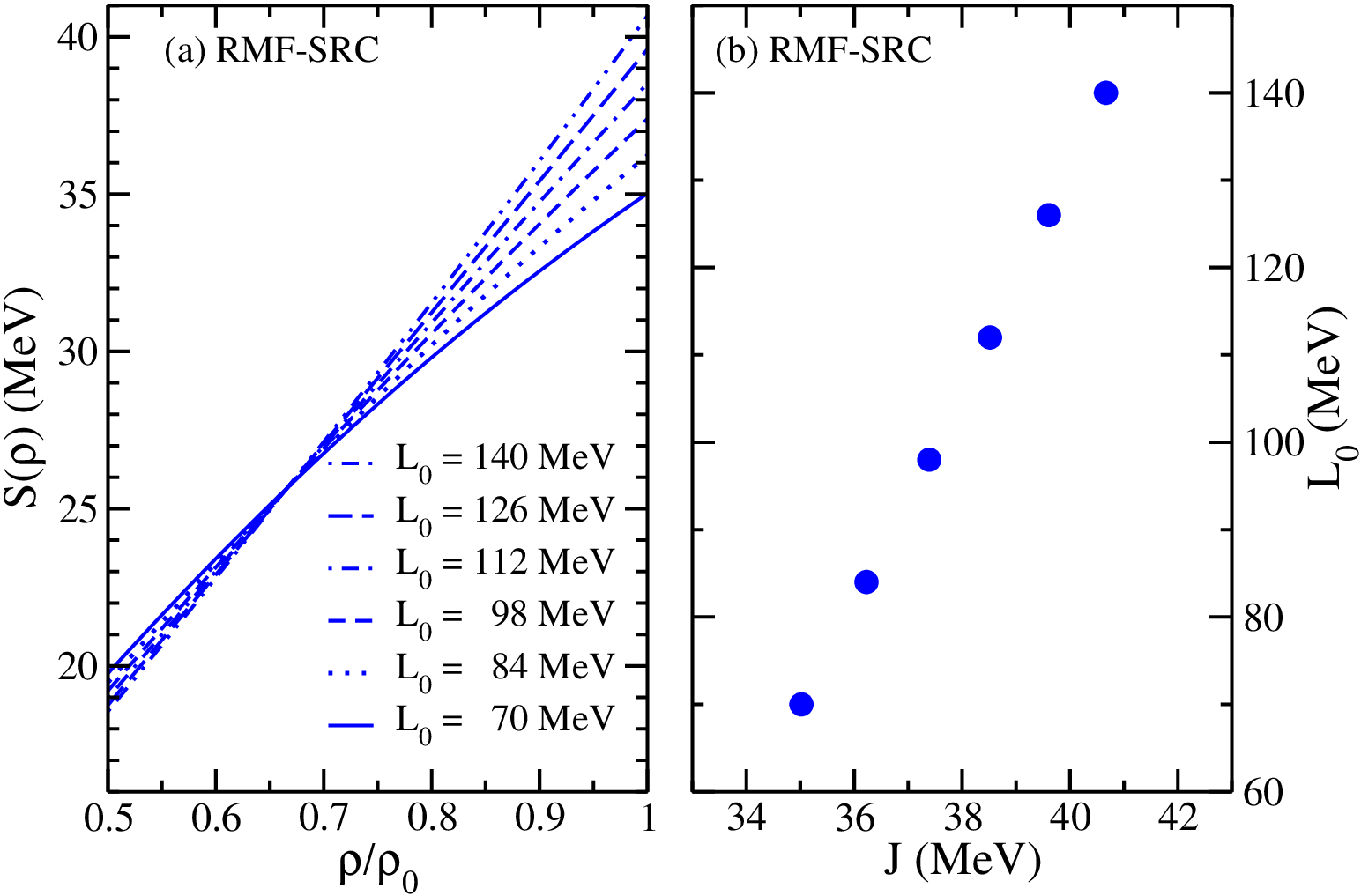}
\caption{(a) Density dependence of the symmetry energy, and (b) linear correlation presented by $L_0$ and~$J$. Results of both panels: parametrizations of the RMF-SRC model (hadronic sector).}
\label{lj}
\end{figure}

Finally, we also investigate the influence of the isovector sector of the relativistic model with SRC and DM. In particular, we verify how the symmetry energy slope affects some quantities related to the neutron stars. In the hadronic matter context, the symmetry energy is defined by $\mathcal{S}(\rho) = \frac{1}{8}\frac{\partial^2(\mathcal{E}/\rho)}{\partial y^2}\bigg|_{\rho,y=1/2}$,
with its slope given by $L = 3\rho\frac{\partial\mathcal{S}}{\partial\rho}$. It is known that the value of the symmetry energy slope at the saturation density, $L_0=L(\rho_0)$, can significantly change some stellar matter properties, such as the neutron star mass-radius profile (see e.g. Ref.~\cite{baoanli}). In order to perform this analysis, we use different parametrizations obtained from the independent variation of $L_0$ with the remaining quantities fixed in the values shown in Table~\ref{tabbulk}, as in the former cases. Before presenting the results, a brief explanation regarding this variation is in order. It is known that some bulk properties can be correlated, and such a relationship is useful to constrain the microphysics related to the equation of state of the hadronic model. For the isovector sector, this is the case of $L_0$ and $J=\mathcal{S}(\rho_0)$, for which many approaches suggest a strong
correlation~\cite{baoanli,drischler,bianca}. In particular, the authors of Ref.~\cite{bianca} shown that a crossing point in the density dependence of $\mathcal{S}(\rho)$ can be seen as a sign of the linear correlation between $L_0$ and~$J$. Here we impose that all parametrizations constructed by the variation of $L_0$ exhibit this crossing point in the $\mathcal{S}(\rho)$ curve, by fixing $\tilde{J}\equiv\mathcal{S}(2\rho_0/3)=25.68$~MeV, namely, the value obtained from the reference parametrization FSU2R. This value is compatible with $\mathcal{S}_1\simeq26$~MeV found in Refs.~\cite{piekaprex2,pieka2001}. In Fig.~\ref{lj} we show the density dependence of $\mathcal{S}(\rho)$ as well as the linear correlation presented by the parametrizations generated by the variation of $L_0$.

\begin{figure}[!htb] 
\centering
\includegraphics[scale=0.364]{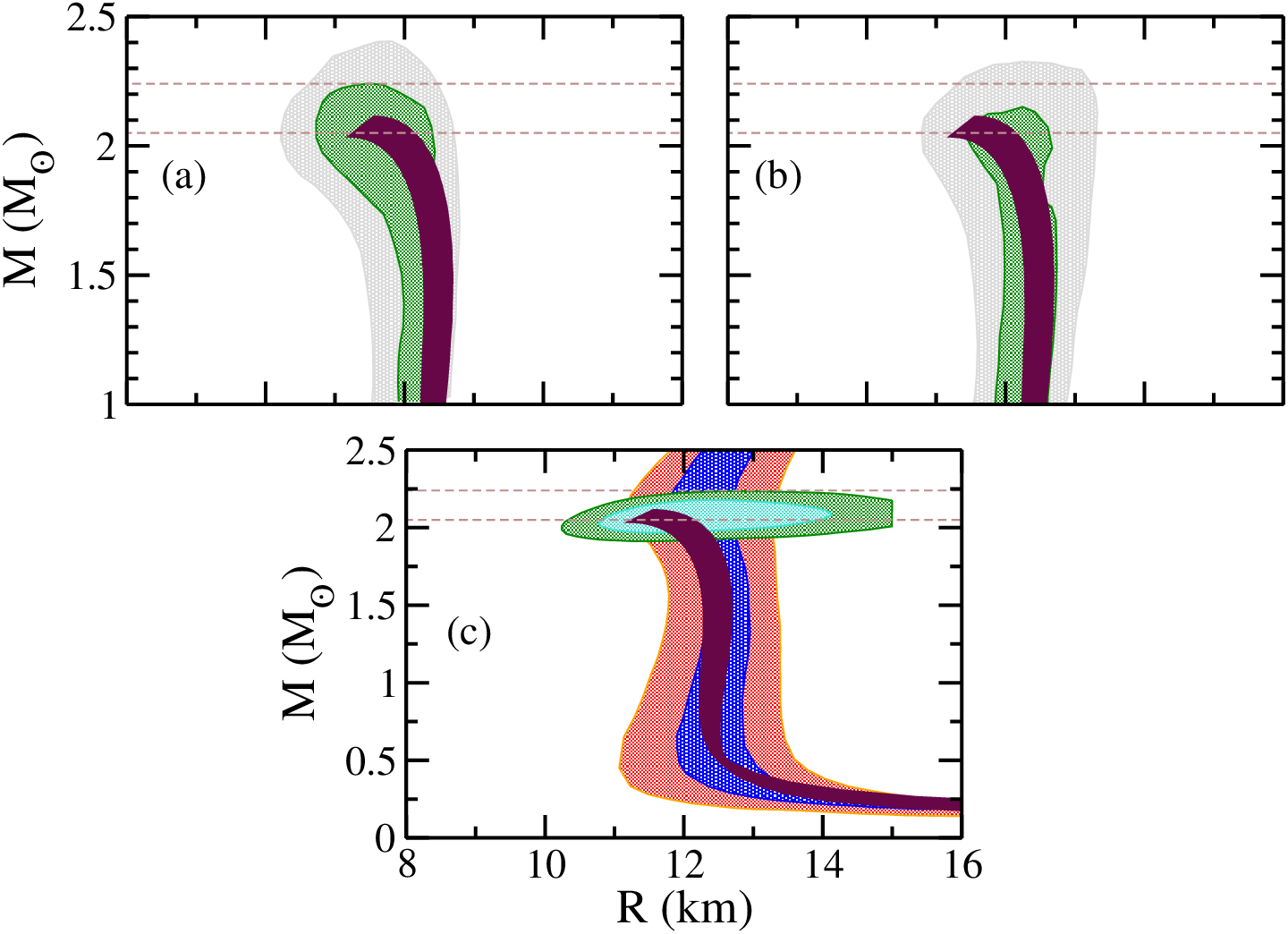}
\caption{Mass-radius diagrams. Brown region: band generated by the variation $L_0=(106\pm37)$~MeV~\cite{piekaprex2} in the \mbox{RMF-SRC} model with DM content ($k_F^{\mbox{\tiny DM}}=0.03$~GeV). Other regions: the same as in Figs.~\ref{mr1} and~\ref{mr2}. Horizontal lines: data taken from Ref.~\cite{cromartie}.}
\label{bandl0}
\end{figure}

The region obtained from the aforementioned variation of $L_0$ is depicted in Fig.~\ref{bandl0}.
In this figure, we take the values of $L_0=(106\pm37)$~MeV~\cite{piekaprex2} claimed to be 
consistent with the updated results provided by the Lead Radius EXperiment~(\mbox{PREX-2}) 
collaboration regarding the neutron skin thickness values of $^{208}\rm Pb$~\cite{prex2}. For this 
case, we verify that is possible to construct parametrizations consistent with the bands determined 
from NICER, LIGO and Virgo, and data from radio observations, and also with the range given by 
$M=2.14^{+0.10}_{-0.09}M_{\odot}$~\cite{cromartie}. Moreover, due to the correlation between $L_0$ 
and $J$, we find the range of $34.9\,\mbox{MeV}\leqslant J \leqslant 40.9~\mbox{MeV}$ for the 
symmetry energy at the saturation density. These values are compatible with the PREX-2 outcome of 
$J=(37.7\pm4.1)$~MeV, inferred from the range of $L_0$~\cite{piekaprex2}, and also with 
$J=(38.1\pm4.7)$~MeV determined in Ref.~\cite{piekaprex2}. Furthermore, we also calculate 
$\tilde{L}\equiv L(2\rho_0/3)$ and find the interval of $65.6\,\mbox{MeV}\leqslant \tilde{L} 
\leqslant 87.6~\mbox{MeV}$, which is completely inside the range presented in 
Ref.~\cite{piekaprex2}, namely, $\tilde{L}=(71.5\pm22.6)$~MeV. As a last comparison regarding $L_0$, 
it is also worth noting that the range $L_0=(106\pm37)$~MeV~\cite{piekaprex2} overlaps with other 
estimations for this isovector bulk parameter obtained from measurements of the spectra of charged 
pions, namely, $42\leqslant L_0\leqslant 117$~MeV~\cite{pions}.

\section{Summary and concluding remarks}
\label{secsummary}

In this work, we analyze the existence of dark matter in the composition of astrophysical systems, 
more specifically neutron stars. In order to corroborate this possibility, we compare our results to 
the recent observational constraints coming from the combined analysis of data from the NICER 
mission, LIGO and Virgo collaborations, and mass measurements from radio 
observations~\cite{arx78,arx79,arx80,arx81}. The DM model used here is composed by the lightest 
neutralino, a candidate that belongs to a class of Weakly Interacting Massive Particles present in 
supersymmetric theories beyond the Standard Model~\cite{cand1,cand2}, with a mass of 
$M_\chi=200$~GeV. We also adopt the Higgs boson as the mediator between neutralino and nuclear 
matter formed by protons and neutrons. The coupling used here between the Higgs and the neutralino 
is compatible with experimental boundaries provided by \mbox{PandaX-II}~\cite{pandaxII}, 
LUX~\cite{lux}, and DarkSide~\cite{darkside} collaborations for the spin-independent scattering 
cross-sections. For the hadronic sector, we use a relativistic mean-field model that takes into 
account the phenomenology of short-range 
correlations~\cite{nature,hen2017,ye2018,Egiyan2006,Frankfurt1993,Fomin2012,Atti2015,Shneor2007,
Tang2003,Li2019,Schmookler2019,Duer2019,Ryck2019,Chen2017}, implemented through the modification of 
the momentum distribution function, replaced by the one predicting a high momentum tail proportional 
to~$1/k^4$~\cite{cai,lucas}.

Through a thermodynamical analysis of the DM coupled to the nuclear matter model, we find that it is possible to describe the whole system only by adding to the original RMF model the kinetic terms related to the neutralino in the energy density and pressure. This feature is basically due to the smallness of the Higgs mean field $h$ as a function of density, which justifies the absence of higher-order terms in the DM Lagrangian density regarding this boson~\cite{rmfdm2}. Therefore, the effective nucleon mass can be expressed as in the original RMF model, namely, $M^*\simeq M_{\mbox{\tiny nuc}}-g_\sigma\sigma$, and the effective neutralino mass reads $M^*_\chi\simeq M_\chi$, i.e, there is no density dependence on that. This fact enables us to write all thermodynamical quantities that are obtained through derivatives of the density exactly with the same expression of the RMF model. This is the case of the chemical potentials, symmetry energy, and symmetry energy slope, for instance. 

By choosing the \mbox{FSU2R-SRC} parametrization in the hadronic sector, we verify that is possible to reach the mass-radius constraints~\cite{arx78,arx79,arx80,arx81}. In this case, we show that the decreasing of the maximum neutron star mass induced by the addition of dark matter, i.e. by growing $k_F^{\mbox{\tiny DM}}$, is balanced by the increase of the same quantity produced by the short-range correlations. We also observe agreement of our results with the boundaries of $M=2.14^{+0.10}_{-0.09} M_{\odot}$ from Ref.~\cite{cromartie}.

Finally, we study the impact on the mass-radius diagram caused by the uncertainties in the bulk parameters related to the hadronic sector, namely, incompressibility, effective nucleon mass, and symmetry energy slope, all of them evaluated at the saturation density. We perform independent variations in $K_0$~($220\,\mbox{MeV}\leqslant K_0 \leqslant 260~\mbox{MeV}$), $m^*$~($0.58\leqslant m^* \leqslant 0.64$) and $L_0$~($69\leqslant L_0\leqslant 143$~MeV) in the \mbox{RMF-SRC} model with dark matter content ($k_F^{\mbox{\tiny DM}}=0.03$~GeV). Our results indicate that the change in $K_0$ generates a very narrow band in comparison to the one constructed through the variation of $m^*$, for instance. In addition, we verify that these two bands do not reach the range of $M=2.14^{+0.10}_{-0.09}M_{\odot}$~\cite{cromartie}. On the other hand, the uncertainty in $L_0$ produces a NS mass-radius band that satisfies this particular constraint as well as the ones determined from recent astrophysical observations. In addition, we emphasize that the uncertainty in $L_0$ is the one in agreement with results reported by the updated \mbox{PREX-2} collaboration on the neutron skin thickness values of $^{208}\rm Pb$~\cite{piekaprex2}.

\section*{ACKNOWLEDGMENTS}
This work is a part of the project INCT-FNA proc. No. 464898/2014-5. It is also supported by Conselho Nacional de Desenvolvimento Cient\'ifico e Tecnol\'ogico (CNPq) under Grants No. 406958/2018-1, 312410/2020-4 (O.L.), No. 433369/2018-3 (M.D.), and 308486/2015-3 (T.F.). We also acknowledge Funda\c{c}\~ao de Amparo \`a Pesquisa do Estado de S\~ao Paulo (FAPESP) under Thematic Project 2017/05660-0 (O.L., M.D., T.F.) and Grant No. 2020/05238-9 (O.L., M.D.).

\end{document}